\documentclass[]{aa}

\usepackage{graphicx}
\usepackage{natbib}
\usepackage{txfonts}
\begin{document}

 \title{Unveiling the near-infrared structure of the massive-young
     stellar object NGC~3603
     IRS\,9A* with sparse aperture masking and spectroastrometry}

   \author{J. Sanchez-Bermudez\inst{\ref{inst1},\ref{inst2},\ref{inst3}} \and
     C. A. Hummel \inst{\ref{inst3}} \and P. Tuthill
     \inst{\ref{inst4}} \and A. Alberdi
     \inst{\ref{inst2}} \and
     R. Sch\"odel \inst{\ref{inst2}} \and S. Lacour \inst{\ref{inst5}}
   \and T. Stanke \inst{\ref{inst3}}}

      \institute{
Max-Planck-Institut f\"ur Astronomie, K\"onigstuhl 17, 69117
Heidelberg, Germany. \email{jsanchez@mpia.de}\label{inst1}
\and
Instituto de Astrof\'isica de Andaluc\'ia (CSIC), Glorieta de la Astronom\'ia S/N, 18008 Granada, Spain. \label{inst2}
         \and
        European Southern Observatory, Karl-Schwarzschild-Stra$\beta$e
        2, 85748 Garching,
        Germany. \label{inst3}
\and
        Sydney Institute for Astronomy, School of Physics, The
        University of Sydney, N. S. W. 2006, Australia. \label{inst4}
\and
        LESIA/Observatorie de Paris, CNRS, UPMC, Universit\'e
        Paris Diderot, 5 place Jules Janssen, 92195 Meudon,
        France. \label{inst5} 
}

\titlerunning{IRS9A Near-infrared structure}
   \date{}
% \abstract{}{}{}{}{} 
% 5 {} token are mandatory
 
  \abstract
  {Contemporary theory holds that massive stars gather mass during their initial phases via
    accreting disk-like structures. However, conclusive evidence for disks has remained 
    elusive for most massive young objects. This is mainly due to significant 
    observational challenges: objects are rare and located at great distances within 
    dusty, highly opaque environments.
    Incisive studies, even targeting individual objects, are therefore relevant to the progression
    of the field. 
    NGC 3603 IRS\,9A* is a young massive stellar object that is still surrounded by an envelope of 
    molecular gas for which previous mid-infrared observations with
    long-baseline interferometry have provided evidence of a plausible disk of 50\,mas diameter at its core. }
  {This work aims at a comprehensive study of the physics and morphology of IRS\,9A 
    at near-infrared wavelengths. }
   {New sparse aperture-masking interferometry data, taken with the
    near-infrared camera NACO of the Very Large Telescope (VLT) at $K_s$ and $L'$
    wavelengths, were analyzed together with archival high-resolution
    H$_2$ and Br$\gamma$ lines obtained with the cryogenic
    high-resolution infrared schelle spectrograph (CRIRES).}
   {The trends in the calibrated visibilities at $K_s$ and $L'$ bands
    suggest the presence of a partially resolved compact object with
    an angular size of $\leq$30 mas at the
     core of IRS\,9A, together with the presence of over-resolved flux.  The
     spectroastrometric signal of the $H_2$ line, obtained from the CRIRES spectra, shows that this spectral feature proceeds from the large-scale extended
   emission ($\sim$ 300 mas), while the Br$\gamma$ line
   appears to be formed at the core of the object ($\sim$ 20 mas). 
To better understand the physics that drive IRS\,9A, we have
  performed continuum radiative
transfer modeling. Our best model
   supports the existence of a compact disk with an
   angular diameter of 20\,mas, together with an outer envelope of 1'' exhibiting a polar cavity 
   with an opening angle of $\sim$30$^{\circ}$. This model reproduces the MIR
   morphology that has previously been derived in the literature and
   also matches the spectral energy distribution of the source.}
{Our observations and modeling of IRS\,9A support the presence of
  a disk at the core, surrounded by an envelope. This scenario is
  consistent with the brightness distribution of the source for near-
  and mid-infrared wavelengths at various spatial scales. However,
  our model suffers from remaining inconsistencies between SED modelling and
  the interferometric data. Moreover, the Br$\gamma$
  spectroastrometric signal indicates that the core of IRS\,9A exhibits some form of 
  complexity such as asymmetries in the disk. 
  Future high-resolution observations are required to confirm the disk/envelope
  model and to flesh out the details of the physical form of the inner regions of
  IRS\,9A.}


   \maketitle
%
%________________________________________________________________
\section{Introduction}

Massive stars play an important role in the evolution of 
galaxies. They exert a tremendous influence
on their surroundings through the mass flows that are produced by their strong
stellar winds and by their deaths in the form of supernova explosions. However,
in spite of their importance, our knowledge about their births and evolution is
still not conclusive. This is mainly because their evolutionary
time-scale is of the order of a few Ma. They are formed in dense
molecular clouds, that are highly opaque and, by the time they reach the
zero-age main sequence, they are still embedded in their parental
cloud \citep{Churchwell_2002}. Hence, the observation of their initial evolutionary phases is
challenging. Furthermore, they are scarce and usually born in dense
clusters that are
located at large distances ($\geq$ 1 kpc), thus the use of
high-angular resolution techniques is required to study them \citep[see the review of ][]{Zinnecker_2007}. 

\begin{figure*}
\centering
\includegraphics[width=15 cm ]{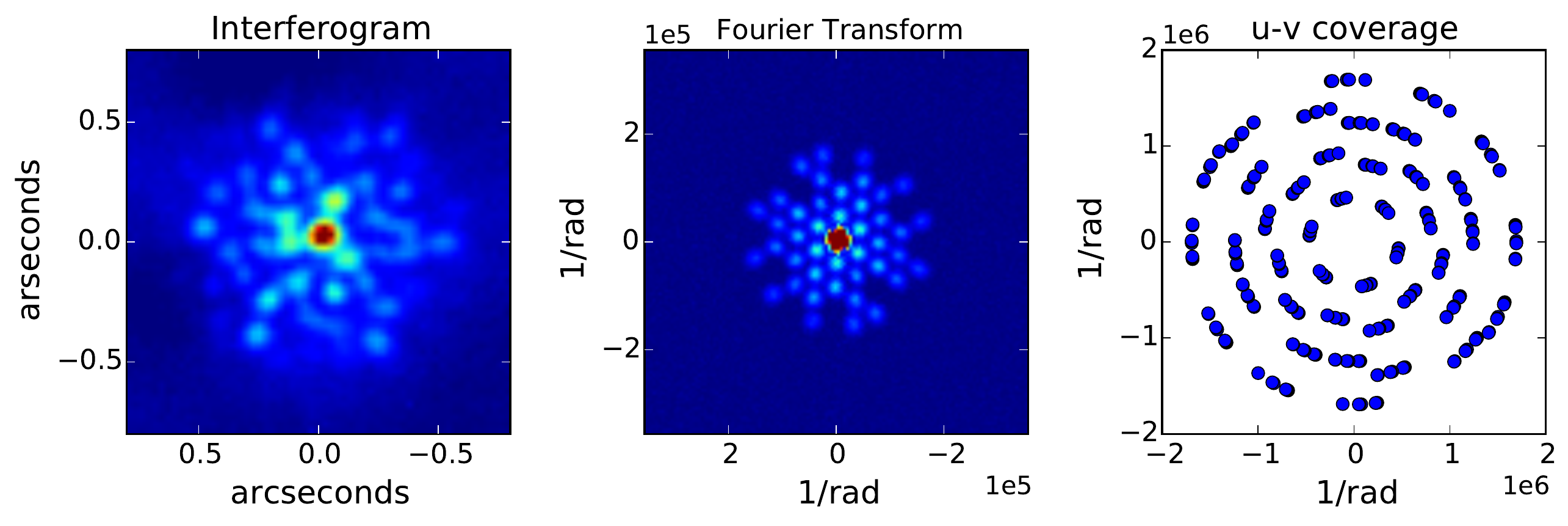} 
\caption[IRS\,9A interferogram]{Left: IRS\,9A: Average interferogram in the image
  plane observed in a sparse aperture masking (SAM) data cube. Center: Fourier
  transform of an average interferogram. The 42 blue spots
  correspond to the sampled visibilities and their complex conjugates,
  while the central red spot is the DC component of the Fourier
  transform and it is proportional to the squared of the total flux in
  the interferogram. Right: UV
  coverage of our SAM observations.} 
 \label{fig:calib}
\end{figure*}

There are two contemporary models that seek to explain high-mass
star formation: core collapse and competitive
accretion. Although both models predict the existence of accretion disks
in massive young stellar objects (MYSOs), there are still important
questions about the
physics of these structures (e.g., sizes, accretion rates, densities, ages). This is mainly because of
the lack of observational evidence of such disks owing to the high
extinction that affects MYSOs and the necessity of high-angular
resolution at infrared wavelengths. One of the most important questions is
the role of radiation pressure from luminous central stars
(L$\sim$10$^4$-10$^5$L$_{\odot}$) on the physics of the accretion
disks.  Massive stars have short Kelvin-Helmholtz ($\sim$ 10$^4$-10$^5$
years) scales, compared to their accretion time-scale. Therefore, by the time the central star is fusing hydrogen,
it is still accreting material. Onset of fusion and its increase
of a star's effective temperature lead to a significant increase of
overall radiative output and UV radiation, which poses strong
constraints on how high-mass stars can accrete matter. For example, Bondi-Hoyle accretion
simulations predict that radiation pressure could stop accretion for
massive stars $\ge$ 10M$_{\odot}$ \citep{Edgar_2004}. These results highlight the
necessity to have protostellar disks to shield the accreting material
from the radiation pressure
to
form the most massive stars.

During the last decade, significant progress has
been made, not only in the study of massive protostellar disks, but
also of the whole morphology of MYSOs. From the theoretical point of view, many studies have been conducted to
explain the different properties of disks in MYSOs, most of them in the
context of core collapse. For example, \citet{Vaidya_2009} performed steady
state models of thin disks around MYSOs to investigate the role of
accretion. Additionally, \citet{Krumholz_2009} performed 3D
simulations of forming massive stars showing the presence of rotational structures (i.e. accretion disks) with high accretion rates ($\sim$10$^{-4}$M$_{\odot}$/yr). These simulations
also support evidence that gravitational and Rayleigh-Taylor
instabilities in the disk-like structures overcome the radiation
pressure of the star and lead to the formation of
companions \citep[but see a different view in ][]{Kuiper_2012}. \citet{Seifried_2011, Seifried_2012_erratum} present
hydrodynamical simulations of collapsing massive cores to explain the
role of magnetic fields on the stability of massive disks, concluding
that magnetic pressure actively contributes to their stability.
On the other hand, \citet{Kuiper_2014} 
determine, through hydrodynamical simulations, that the accretion time of the disk is strongly correlated
with the time at which protostellar outflows are present. 

Alternatively, significant observational
progress was triggered by the advent of adaptive optics (AO) at
8-10m class telescope, infrared interferometry, as well as by the improved sensitivity and
angular resolution of millimeter interferometers. For example, radio
interferometric observations have found evidence of molecular
outflows and jets launched at the core of
the MYSOs
\citep[e.g., ][]{Beuther_2002, Beuther_2005}, as well as evidence of
disk-like structures that are perpendicular to such outflows \citep[e.g., ][]{Beltran_2005}, thus providing us with a picture
consistent with the predicted circumstellar disks. Recently, \citet{Boley_2013} performed a survey of disks
around a sample of 24 intermediate and high-mass stellar
objects at MIR wavelengths with MIDI at the Very Large
Telescope Interferometer (VLTI). Almost all
the observed massive stars of their sample presented elongated
structures at scales $\le$100 AU, which could be associated with disks
and/or outflows. Additional MIR interferometry case studies on
individual objects include: M8E-IR \citep{Linz_2009}, M17 SW IRS1
\citep{Follert_2010}, AFGL 4176 \citep{Boley_2012}, CRL~2136 \citep{de_Wit_2011}, and NGC~2264~IRS~1*
\citep{Grellmann_2011}. We also note that the work of \citet{de_Wit_2007} was one of the first attempts to
perform simultaneous model-fitting to the spectral energy
distribution (SED) and mid-infrared interferometric visibilities of W33A. At near-infrared wavelengths, perhaps the most
representative case study of disks in MYSOs was performed by
\citet{Kraus_2010}. 
These authors resolved a hot, compact disk around the MYSO IRAS 13481-6124 (M$\sim$20
M$_{\odot}$), carrying out interferometric observations with AMBER/VLTI. The observed
structure exhibits a projected elongated shape of 13x19 AU with a
dust-free inner gap of 9.5 AU, matching the expected location of the
dust sublimation radius for this object. 

Apart from the mentioned examples of observing (or modeling) the
morphology of MYSOs directly, there have been several studies to
characterize MYSOs from their spectra. Two methods are worth mentioning:
(i) the characterization of their SED, and (ii) the analysis of the
flux centroid of absorption or emission lines through
spectroastrometry. An example of the first method is the work of
\citet{Robitaille_2006}. These authors computed a grid of $\sim$10$^5$ SED models of
YSOs up to masses of 50M$_{\odot}$. These models, based on the
  prescription of \citet{Whitney_2003}, include the presence of disks, envelopes, and
cavities to reproduce observed SEDs. \citet{Offner_2012} performed a study to explore the
reliability of SED-fitting to obtain physical parameters of low-mass
YSOs. They find that solely SED-fitting is not enough to fully
characterize the morphology of low-mass YSOs. This is also expected to
be true in the high-mass case. Combining spectroastrometry with
CRIRES and interferometry with AMBER, \citet{Wheelwright_2012} were
able to characterize the morphology of the circumstelar disk around
the binary Be star HD~327083 \citep[see also
][]{Wheelwright_2010}.

\begin{figure*}[htp]
\centering
\includegraphics[width=13 cm]{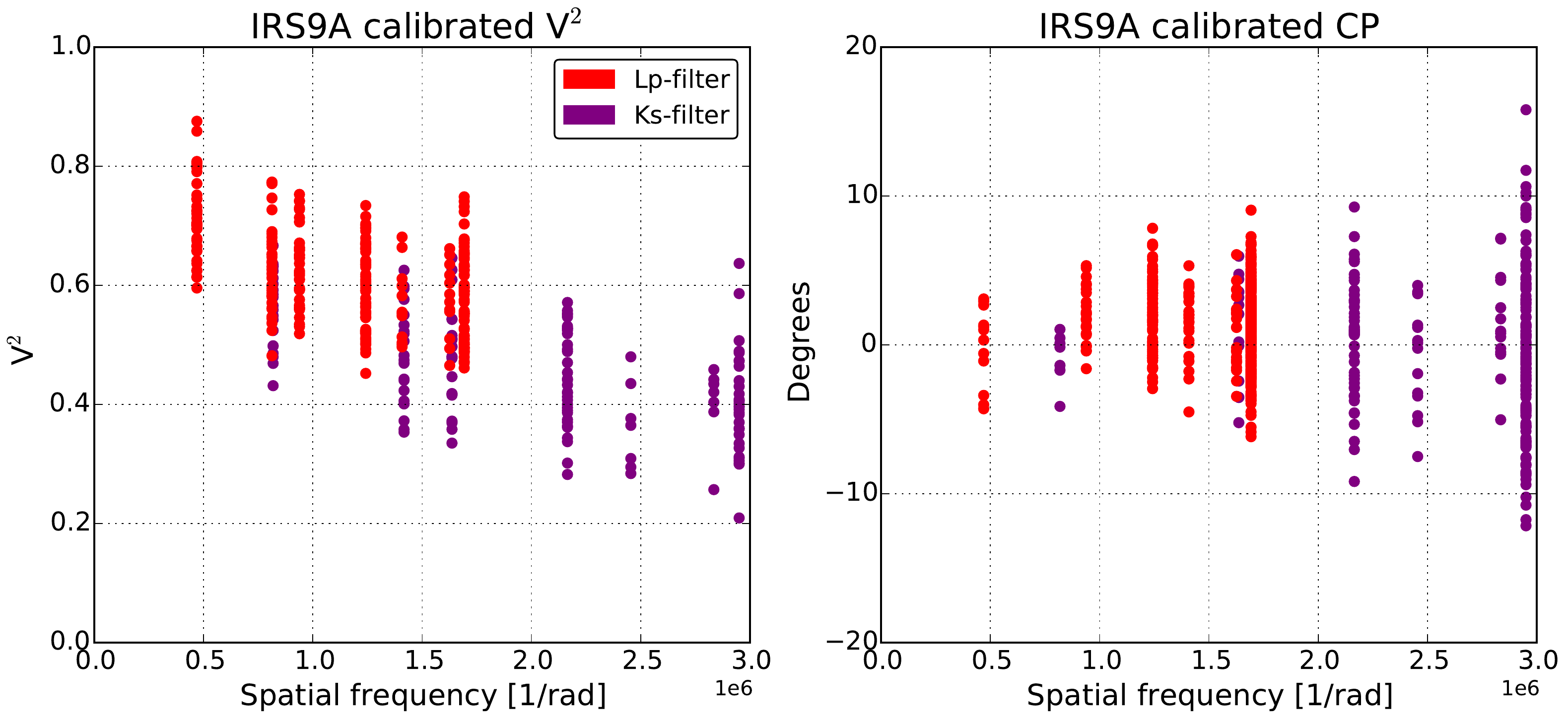} 
\caption[IRS\,9A aperture masking data]{Calibrated squared visibilities (left) and closure phases (right) of
  IRS\,9A. The $V^2$ values imply the presence of a partially resolved
  structure while the CPs are consistent with inversion symmetry of the target.}
 \label{fig:v2_calib}
\end{figure*} 
 
In this work, we explore the morphology of the MYSO NGC\,3603\,IRS\,9A*
(abbreviated hereafter as IRS\,9A) with near-infrared interferometric
observations and long-slit spectra. The target,
which is located in the giant HII region NGC~3603
at a distance of 7$\pm$1 kpc, is a MYSO
with an estimated mass of $\sim$ 40 $M_{\odot}$ \citep{Nurnberger_2003}. Owing to its location in NGC~3603,  the natal cloud of this
luminous source ($\sim$ 2.3 x 10$^5 L_{\odot}$) has been partially eroded by the
stellar radiation of a massive cluster of O and B stars that are located at a
projected distance of 
2.5\,pc to the North-West from the source, thus making it observable at infrared wavelengths. However, its spectral index
($\alpha_{2.2-10\mu m}$=1.37) suggests that the source is still surrounded by
considerable circumstellar material. In fact, the observed
 excess of mid-infrared (MIR) emission, and its positive spectral
 index, resemble the properties of  low-mass class~I YSO
 \citep{Lada_1987, Whitney_2003}. This hypothesis of a partially wind-stripped young object offers the intriguing
 possibility of peering behind the veil at a MYSO during one of its early evolutionary
 stages.

Previous MIR high angular resolution observations of the IRS\,9A morphology
identified at least two components that are associated with a warm-inner
disk-like structure and a cold elongated envelope with
temperatures of $\sim$1000 and $\sim$140 K, respectively
\citep{Vehoff_2010}.  On the one hand, the envelope, resolved with MIR sparse aperture
masking (SAM) observations using the T-ReCS IR camera on Gemini South, exhibits an angular size of 330$\times$280 mas. On the other hand, observations with the ESO MIDI instrument, attached to the
Very Large Telescope
Interferometer (VLTI), implied an angular extension of $\leq50$ mas
for the compact structure.  

Here, we make use of SAM data taken with the ESO near-infrared (NIR)
facility NACO, at the ESO Very Large Telescope (VLT), to study the
morphology at the central region of IRS\,9A. These data are complemented with NIR long-slit spectroscopic
observations with CRIRES/VLT from the ESO archive. We also reanalyze the existing
MIDI/VLTI and T-ReCS/Gemini data, including them in our analysis. The
structure of this work is as follows: Section\,\ref{sec:Observations} describes the observations and data
reduction; our analysis and results for the different used techniques,
as well as our models, are presented in
Section\,\ref{sec:Analysis}; in Section\,\ref{sec:Discussion},
we discuss our results. Finally, in Section\,\ref{sec:conclusions} we present our conclusions. 
%__________________________________________________________________

\begin{table}[htp]
\centering
\caption[Log of the IRS\,9A NaCo/SAM observations]{Observing log of VLT/NaCo-SAM imaging observations taken on
  March 12, 2012}
\begin{tabular}{llllll}
\hline
\hline
Time (UT) & Source & Filter & NDIT$^{\mathrm{a}}$  & DIT$^{\mathrm{b}}$ & Camera\\
\hline
05h 05m & IRS 9A & $K_{s}$ & 8.0 & 10.0  & S27\\
05h 20m & HD 98194$^{\mathrm{c}}$ & $K_{s}$ & 8.0  & 10.0 & S27 \\
05h 32m & IRS 9A & $K_{s}$ & 8.0 & 10.0 & S27 \\
05h 46m & HD 98133$^{\mathrm{c}}$ & $K_{s}$ & 8.0 & 10.0 &S27 \\
06h 00m & IRS 9A & $K_{s}$ & 8.0 & 10.0 & S27 \\
06h 15m & HD 98194$^{\mathrm{c}}$ & $K_{s}$ & 8.0 & 10.0 & S27 \\
06h 27m & IRS 9A & $K_{s}$ & 8.0 & 10.0 & S27 \\
06h 39m & HD 98133$^{\mathrm{c}}$ & $K_{s}$ & 8.0 & 10.0 & S27 \\
07h 10m & IRS 9A & $L'$ & 126 & 0.5  & L27\\
07h 22m & HD 97398$^{\mathrm{c}}$ & $L'$ & 126  & 0.5 & L27 \\
07h 30m & IRS 9A & $L'$ & 126 & 0.5 & L27 \\
07h 41m & HD 98133$^{\mathrm{c}}$ & $L'$ & 126 & 0.5 &L27 \\
07h 50m & IRS 9 A & $L'$ & 126 & 0.5 & L27 \\
08h 02m & HD 97398$^{\mathrm{c}}$ & $L'$ & 126 & 0.5 & L27 \\
08h 13m & IRS 9A & $L'$ & 126 & 0.5 & L27 \\
08h 27m & HD 98133$^{\mathrm{c}}$ & $L'$ & 126 & 0.5 & L27 \\
\hline
\end{tabular}
\begin{list} {}{} \itemsep1pt \parskip0pt \parsep0pt \footnotesize
\item[$^{\mathrm{a}}$] Number of exposures.
\item[$^{\mathrm{b}}$] Detector integration time in seconds. The total integration time of each observation amounts to N$\times$DIT$\times$NDIT.
\item[$^{\mathrm{c}}$] Stars used as calibrators of the sparse
  aperture masking data.
\end{list}
\label{tab:data_NACO} 
 \end{table}

\section{Observations and data reduction}\label{sec:Observations}

\subsection{Sparse aperture-masking data.}

We performed Adaptive Optics (AO) observations of IRS\,9A with NACO in its aperture-masking
mode using  the $K_s$ ($2.2\mu m$), $L'$ ($3.8\mu m$), and $M'$ ($4.7\mu m$) bandpasses
on March 12th, 2012\footnote{ESO programme: 088.C-0093(A)} (UTC),
combined with the S27/L27 cameras (0.027''/pixel) and the 7holes mask. This mask and configuration are 
appropriate for use with targets in the magnitude range of IRS~9A.
The relatively uniform $u-v$ coverage afforded by this mask yielded a synthesized beam
with angular resolution ($\theta$) at each band of  35, 60,
and 80 mas for the $K_s$, $L'$, and $M'$ filters, respectively. 

Observations were taken using repeated standard
calibrator-science target sequences over a period of $\sim$2 hours per filter
(see Table\,\ref{tab:data_NACO}). Each observation on target and
calibrator was composed of four dither positions. Owing to the elevated noise and
the presence of vertical stripes in the upper half of the NACO
detector at the time of the observations, the dither positions were performed following a squared
box in the lower
half of the detector. The repeated observation helped to improve our $u-v$ coverage through Earth
rotation synthesis (see Fig\,\ref{fig:calib}). All SAM data were initially sky subtracted, flat fielded, and bad
pixel corrected. For each different position, the sky frames were constructed via separate
observations of an empty field close to the target. The
sky subtraction did not result in a fully flat background as expected:
on the contrary, patterns were found to remain on the images in the $L'-$filter. The size of these
patterns was larger than the size of the interferogram. 
Because fringes are encoded at high spatial frequency, we assumed that
the effects of these residuals were not significant. Additionally,
because of high airmass and variable seeing, the NACO/SAM $M'-$filter observations had low
photon counts on target, and no interferograms were distinguishable for
almost all the recorded data. Therefore the $M'-$filter was discarded from our analysis.

To improve the signal-to-noise ratio (S/N), a frame-selection 
algorithm rejected approximately 50\% of the recorded images based on two flux
criteria: (i) the total counts on a circular mask of the size of the interferogram, and (ii) the counts at the
peak of the interferogram. The final frames were cropped and stored in
cubes of 128x128 pixels that were centered on the source peak pixel. 
Point-source reference stars located not further than 2$^{\circ}$ from IRS 9A were observed to calibrate
the telescope-atmosphere transfer function. 
Table\,\ref{tab:data_NACO} lists the data sets recorded for both
target and calibrators at each observed filter. 
Figure\,\ref{fig:calib} shows, as an example, the $L'-$filter data
that gives the
average interference pattern over a data cube, the power spectrum,
and the final $u-v$ coverage built through Earth-rotation aperture synthesis.

To reduce the raw data to calibrated squared visibilities ($V^2$) and closure phases (CPs), an
analysis pipeline written in interactive data language (IDL),
originally developed at Sydney University, was employed. 
This code constructs a sampling template (specific to the given configuration of mask and filter) 
to extract complex visibilities from the Fourier transform domain.
Raw interferometric observables 
are obtained as an average over the data cube for all baselines that were sampled by the mask. 
Finally, the calibrated amplitudes are obtained through the ratio between raw $V^2$
on source and calibrator, while calibrated CPs are obtained by
subtracting the CPs of the calibrator from the ones on the target. A more detailed
description of the SAM technique can be found in \citet{Tuthill_2000, Tuthill_2010}. 
Additionally, our data were also reduced with the YORICK pipeline for aperture 
masking \citep[SAMP;][]{Lacour_2011}, which was developed at Paris Observatory. Similar results were obtained using both codes.

To test the level of confidence of both the CPs and the $V^2$ of our
calibrators, we performed a cross-calibration between the
different PSF reference stars to determine the response of the interferometric observables and identify systematics. 
We performed an auto-calibration test for pairs of calibrators taken in the
same quadrant of the detector to minimize the difference in the pixel
gain of the interferogram and, hence, reduce the variability in the $V^2$ level. Our tests confirm that the CPs of the
calibrator have individual standard deviations of
$\sigma_{CPs} \sim 2^{\circ}$.

Figure \ref{fig:v2_calib} presents the calibrated $V^2$ and $CPs$ of our IRS 9A
 $K_s$ and $L'$ observations. The interferometric observables depict a partially
resolved target with point-like symmetry for all the different sampled position
angles. The $V^{2}$ levels vary between $\sim$0.4-0.8 and $\sim$0.2-0.7 for $L'$
and $K_s$, respectively. CPs vary within about -10$^{\circ}$ and
10$^{\circ}$ for the two filters.  

\subsection{CRIRES data}

IRS\,9A was observed with the CRIRES spectrograph \citep{Kaeufl_2004} as part of the 080.C-0873(A) observing
programme. We obtained these data from the ESO archive. Observations were taken using a grating of 31.6 lines/mm centered on two
near-IR emission lines: H$_2$ (2.121 $\mu$m) and Br$_{\gamma}$ (2.166
$\mu$m). The slit length was of $\sim$40'' with a
plate scale of 0.086''/pixel. The covered dispersion ranges were:
2.108-2.150 and 2.161-2.200 $\mu$m for H$_2$ and Br$_{\gamma}$,
respectively. These observations were conducted using a
long-slit with a width of 0.6'', and a resolving power of
  R$\sim$33000 or 9 km/s. The CRIRES observations were not AO-assisted. To obtain information on the emission lines at different orientations
of the target, three position angles (0$^{\circ}$, 90$^{\circ}$, and 128$^{\circ}$)
were covered by the slit for each one of the observed lines. The
exposure time was 60\,s for both observed lines. Two nodding
positions, interwoven with sky observations, were taken at each slit
position. 

The data-reduction was performed using IRAF and proprietary IDL
routines.  First, the 2D spectrum was corrected for the flat-field,
bad pixels, and sky contamination. No strong telluric OH lines were observed at
positions threatening to contaminate astrophysically useful lines in the spectra. 
However, to avoid spurious signals in our data resulting from OH lines, we subtracted
the sky observations from the science spectra. After the previous
correction, the two nodding
positions were aligned along the spatial axis and combined into a single
spectrum. To correct for blending along the dispersion axis, a
2nd-degree polynomial was fit to the stellar
continuum spectrum, which was then subtracted from the
science spectra frames. Figure\,\ref{fig:sa_plots} displays the integrated H$_2$ and
Br$_{\gamma}$ emission lines using an aperture of twice the full-width
at half-maximum (FWHM) of
the continuum emission after processing, as described. 

\begin{figure}[htp]
\centering
\includegraphics[width=8 cm]{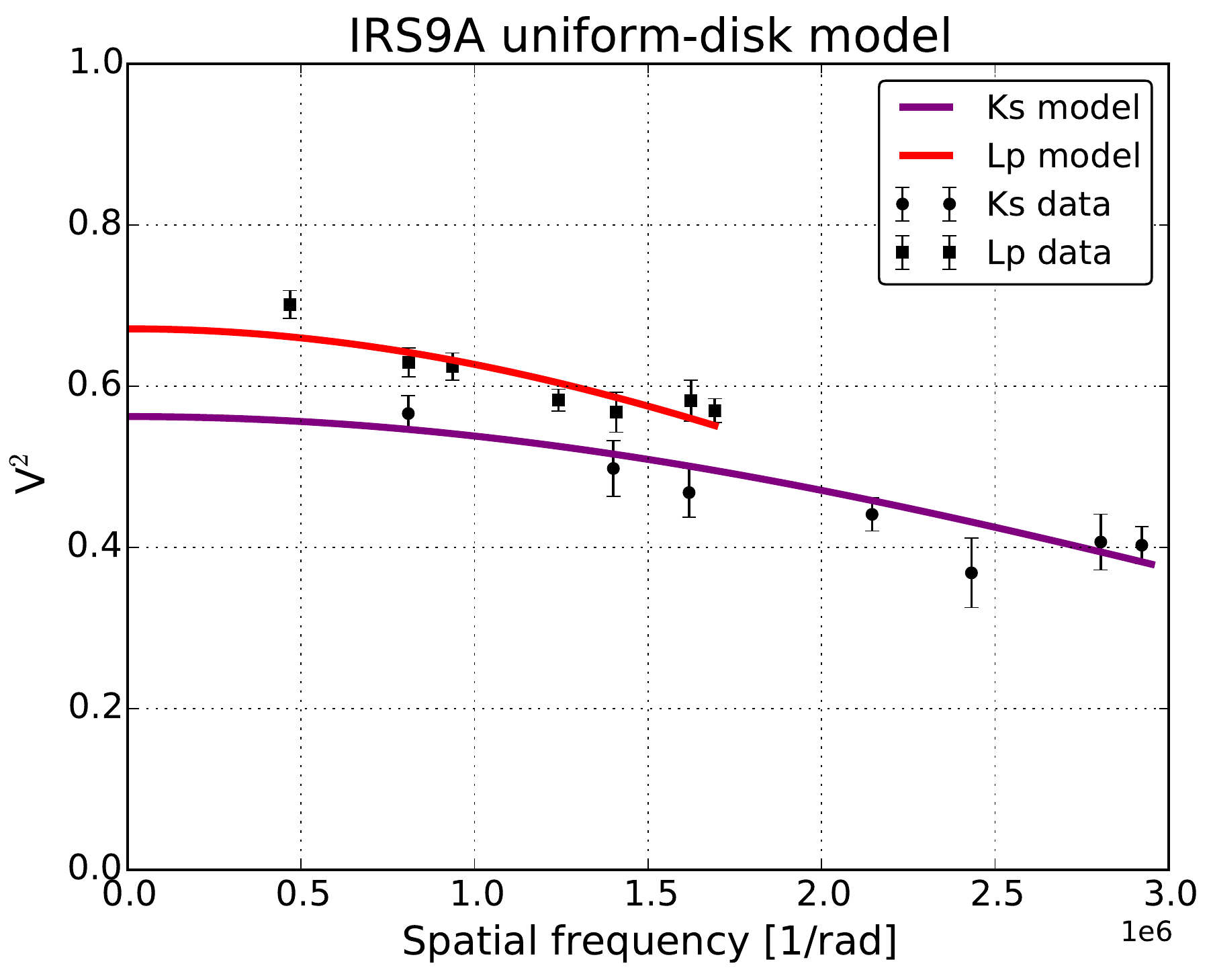} 
\caption[Uniform-disk model applied to the IRS\,9A V$^2$]{Uniform disk model. The mean-weighted $V^{2}$ data points are
  represented by black circles and squares for the $Ks$ and $L'$
  filters, respectively. The red and purple lines represent the
  best-fit model for the $V^{2}$ in $L'$ and $Ks$, respectively.}
 \label{fig:UD_models}
\end{figure}

\section{Analysis and results \label{sec:Analysis}}

\subsection{The core of IRS\,9A probed by K and L-band SAM data \label{sec:NACO_SAM}}

To obtain an estimate of the physical size of the circumstellar
structure around the core of IRS\,9A, we first fit a geometrical 
model of a uniform disk to the $V^{2}$ functions 
from our SAM data. Since we do not completely resolve the compact
structure, we omitted the effect of the inclination in the
model at this stage. The angular size of the disk obtained with our model is,
therefore, an upper limit of the real angular size of the target.
The model of a uniform disk in the Fourier space is given by
the following expression:

\begin{figure*}[htp]
\centering
\includegraphics[width=12 cm]{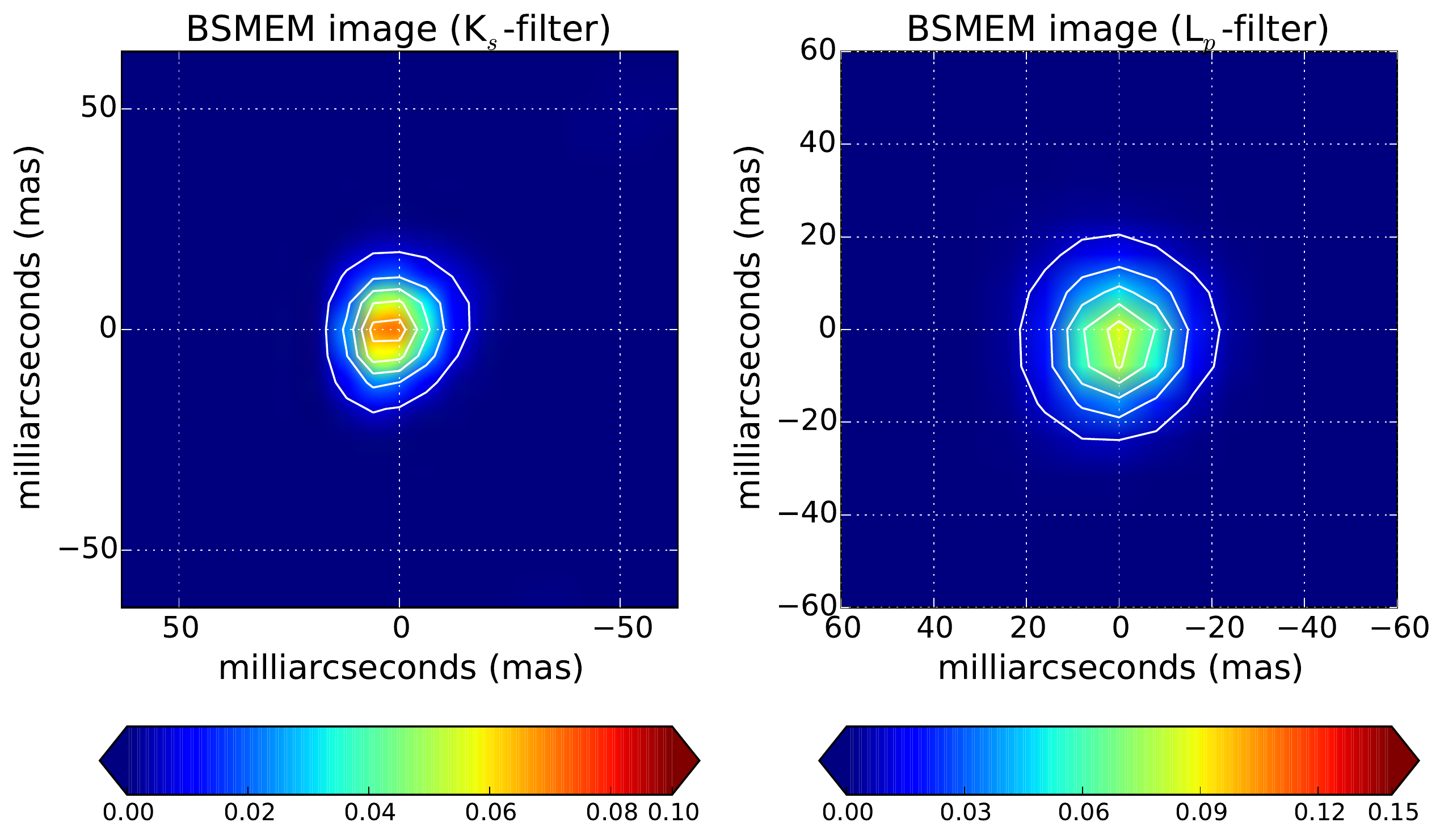} 
\caption[Maximum-entropy images of IRS\,9A]{Reconstructed images of
  IRS\,9A from the NACO/SAM data using BSMEM. The \textit{left} panel
  shows the image for $K_s$ and the \textit{right} panel the image for
$L'$. The total flux of the images is normalized to unity. Each panel has overplotted
black contours that correspond to 10, 30, 50, 70, 90\% of the peak flux.}
 \label{fig:BSMEM}
\end{figure*}

\begin{equation}
V(u,v)^2=\left[\frac{A_1}{A_1+A_2} \right] \left[ 2 \frac{J_1(\pi \theta r)}{\pi \theta r } \right]^2
\,,
\end{equation}

where $J_1$ is a first order Bessel function, $\theta$ is the diameter of the
disk in radians, and $r$ corresponds to the measured spatial
frequencies. The coefficient $A_1$ corresponds to the squared correlated flux
at zero baseline for the disk, and $A_2$ to the squared uncorrelated flux,
where $A_1$+$A_2$=1. 
To optimize the model fit, we used a Levenberg-Marquardt algorithm
that was
implemented at the IDL $MPFIT$ package \citep{Markwardt_2009}. 
Figure\,\ref{fig:UD_models} displays the best-fit model obtained for both
filters, while Table\,\ref{tab:UD_V2} gives values of the best-fit
parameters and their uncertainties.  The fact that $A_1$ does not
reach unity in both bands provides evidence for 
over-resolved extended flux. \citet{Nurnberger_2008} shows the
$L'-$band large scale structure of the circumstellar environment
around IRS\,9A. That work presents an East-West elongated
emission around IRS\,9A with an angular size of around
$\sim$1.0-1.5'' (or $\sim$7x10$^3$-10x10$^3$ AU at a distance of 7kpc). The angular extension of that
diffuse emission is larger than the angular resolution of our shortest
baseline ($\theta$=0.5''), consequently, this can explain the origin of the over-resolved flux
observed at our NACO/SAM data at $L'$. In the case of $K_s$, the presence of circumstellar
matter or a halo of scattered light could also explain the over-resolved
flux that is observed in our interferometric data. This finding will be addressed in subsequent sections. 

As well as using the geometrical model, we performed image
reconstruction with the calibrated data sets. The maximum entropy code
BSMEM was used for this purpose
\citep{Buscher_1994}. As a prior for the reconstruction, we used the
assumption of a uniform disk. The initial model setup included a disk with a diameter
of 50 and 100 mas for $K_S$ and $L'$, respectively. The final images were obtained after 13 and 40 iterations. Figure\,\ref{fig:BSMEM}
displays the reconstructed images. As was expected, IRS\,9A
appears quite compact, however a slight
elongation is seen in the East-West direction in the $K_S-$filter. On the
other hand, at $L'$, the source looks quite symmetrical. We note that
maximum-entropy (MEM) images usually have
a super-resolution beyond the diffraction limit, since the
reconstruction depends only on the entropy statistics \citep[for a
more detailed description, see][]{Monnier_2003, Monnier_2012}. Images
presented in Fig\,\ref{fig:BSMEM} are just the result of the
reconstruction and have not been subsequently convolved with any
beam to make it more evident that the source does not resemble a point-like object
(although it is not completely resolved). 

\begin{table}[ht]
  \caption[Parameters of best-fit uniform-disk model to the $V^{2}$
  IRS\,9A data ]{Parameters of best-fit uniform-disk model to the $V^{2}$ data }
\label{tab:UD_V2}
\centering
\begin{tabular}{l c c c}
\hline\hline
Filter & Parameter & Value & $\sigma^{\mathrm{a}}$\\
\hline
$K_s$ & $A_1^{\mathrm{b}}$ & 0.56 & 0.02 \\
   & D [mas]$^{\mathrm{c}}$ & 27.5 & 2.2 \\
\hline   
$L'$ & $A_1^{\mathrm{b}}$ & 0.67 & 0.02 \\
   & D [mas]$^{\mathrm{c}}$ & 34.2 & 4.5 \\
\hline
\end{tabular}
\begin{list} {}{} \itemsep1pt \parskip0pt \parsep0pt \footnotesize
\item[$^{\mathrm{a}}$] 1-$\sigma$ errors of the best-fit
  parameters, assuming a reduced $\chi^2$ of one.
\item[$^{\mathrm{b}}$] $V^2$ value at zero-baseline.
\item[$^{\mathrm{c}}$] Diameter of the fitted-disk model in units of milliarcseconds.
\end{list}
\end{table}

\subsection{The H$_2$ and Br$\gamma$ spectroastrometric signals \label{sec:spectro}}

To determine and characterize the region from which H$_2$ and Br$\gamma$ emission lines arise, we
extracted the spectroastrometric (SA) signal from both lines using custom
IDL/Python routines. We note that the recommended procedure to calibrate the spectroastrometric signals requires observations with parallel and anti-parallel slit position angles \citep{Brannigan_2006}. Since the archival CRIRES data lack anti-parallel position angles, we limit our analysis to the following: Spectroastrometry tracks the centroid of the stellar continuum and of the
emission line as a function of wavelength, with an astrometric
precision of the order of few milliarcseconds \citep[see
][]{Whelan_2008_SA}. The SA signal was measured on the
non-continuum-subtracted data, and recovered by applying a
Gaussian fitting to the intensity profile of the observed lines along the
spatial axis for each one of the dispersion axis bins. 

To eliminate the centroid shift of the SA signal caused
by the superposition of the stellar continuum, at the emission line position, we
corrected our signal as follows \citep[see][]{Whelan_2008_SA, Pontoppidan_2008_SA}: 

1) The flux of the line ($F_{\lambda (line)}$)  was weighted by the
sum of the flux of the line and continuum ($F_{\lambda (continuum)}$) according to the following
expression:
\begin{equation}
w_{f}=\left[1+ \frac{F_{\lambda (continuum)}}{F_{\lambda (line)}}\right].
%w_{f}=[F_{\lambda (line)}[F_{\lambda (line)}+F_{\lambda (continuum)}]
\end{equation}

2) We computed $w_{f}$ using the average value of $F_{\lambda (line)}$
and $F_{\lambda (continuum)}$. The differences in the centroid position between line and continuum
were thus multiplied by the computed
flux weight, $w_{f}$.

\begin{figure*}[htp]
\centering
\begin{minipage}{\textwidth}
\begin{minipage}[t]{0.32\textwidth}
\includegraphics[width=5.5 cm, height=6.5 cm]{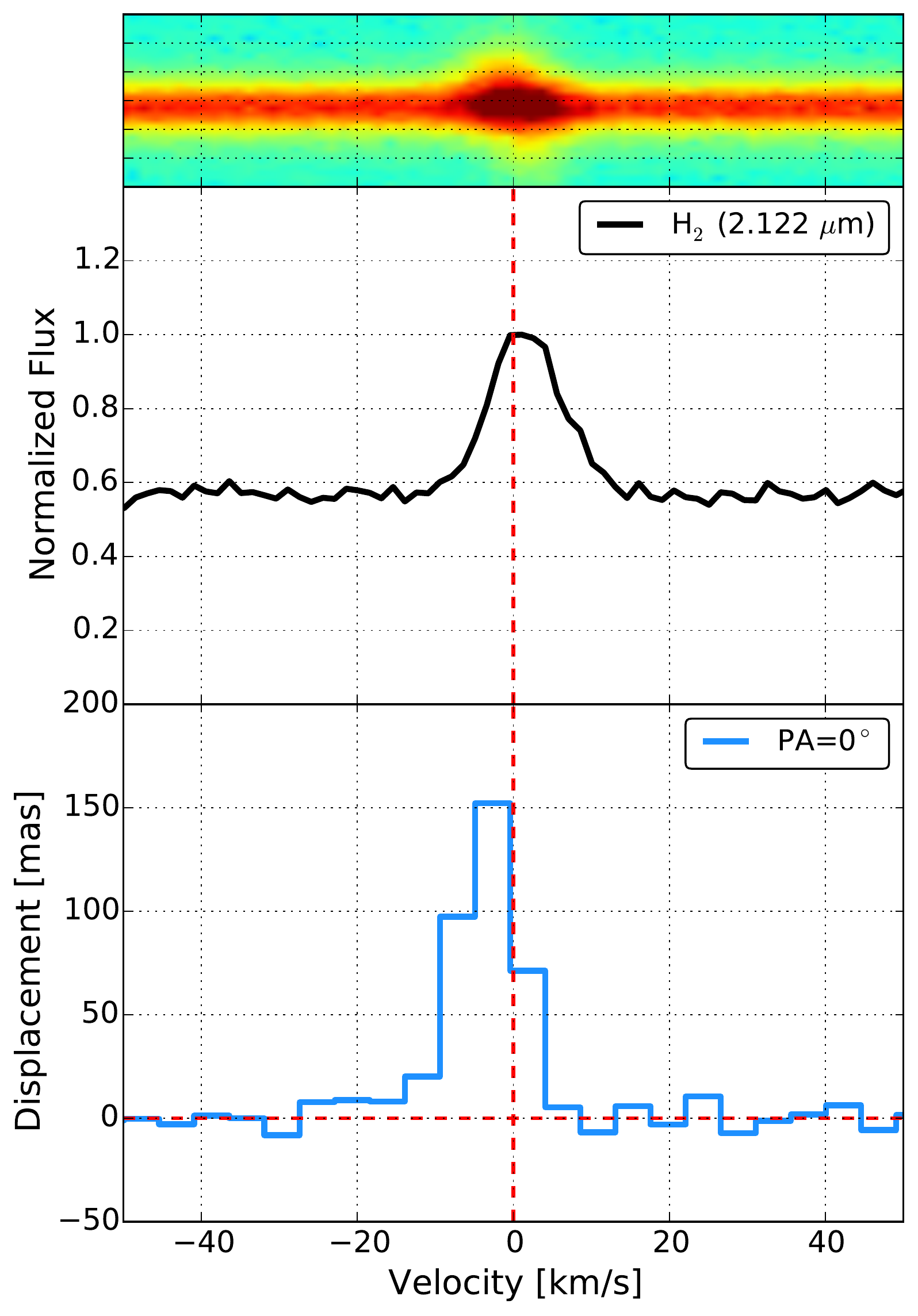} 
\end{minipage}
\begin{minipage}[t]{0.32\textwidth}
\includegraphics[width=5.5 cm, height=6.5 cm]{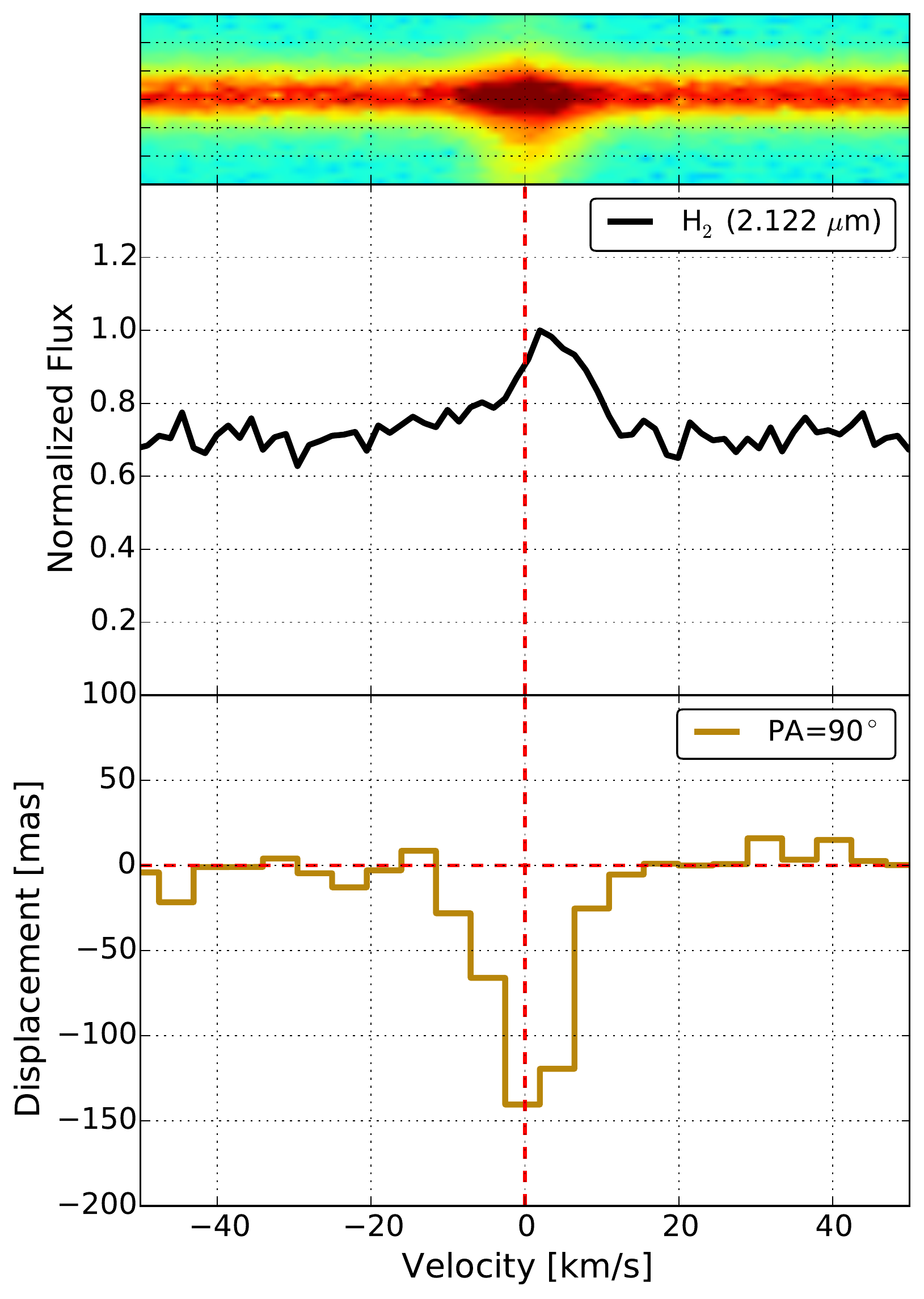} 
\end{minipage}
\begin{minipage}[t]{0.32\textwidth}
\includegraphics[width=5.5 cm, height=6.5 cm]{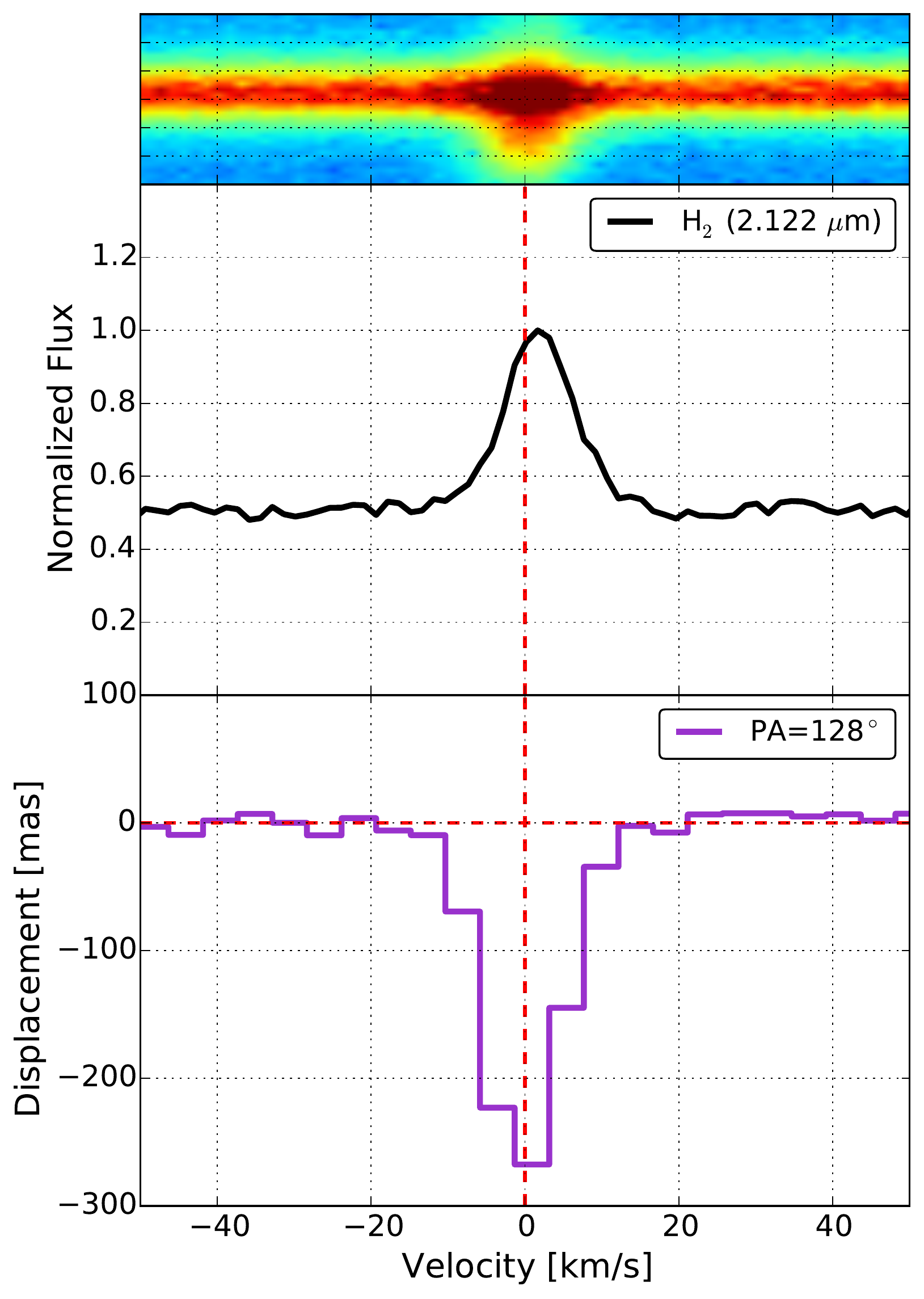} 
\end{minipage}
\end{minipage}

\begin{minipage}{\textwidth}
\begin{minipage}[b]{0.32\textwidth}
\includegraphics[width=5.7 cm, height=6.5 cm]{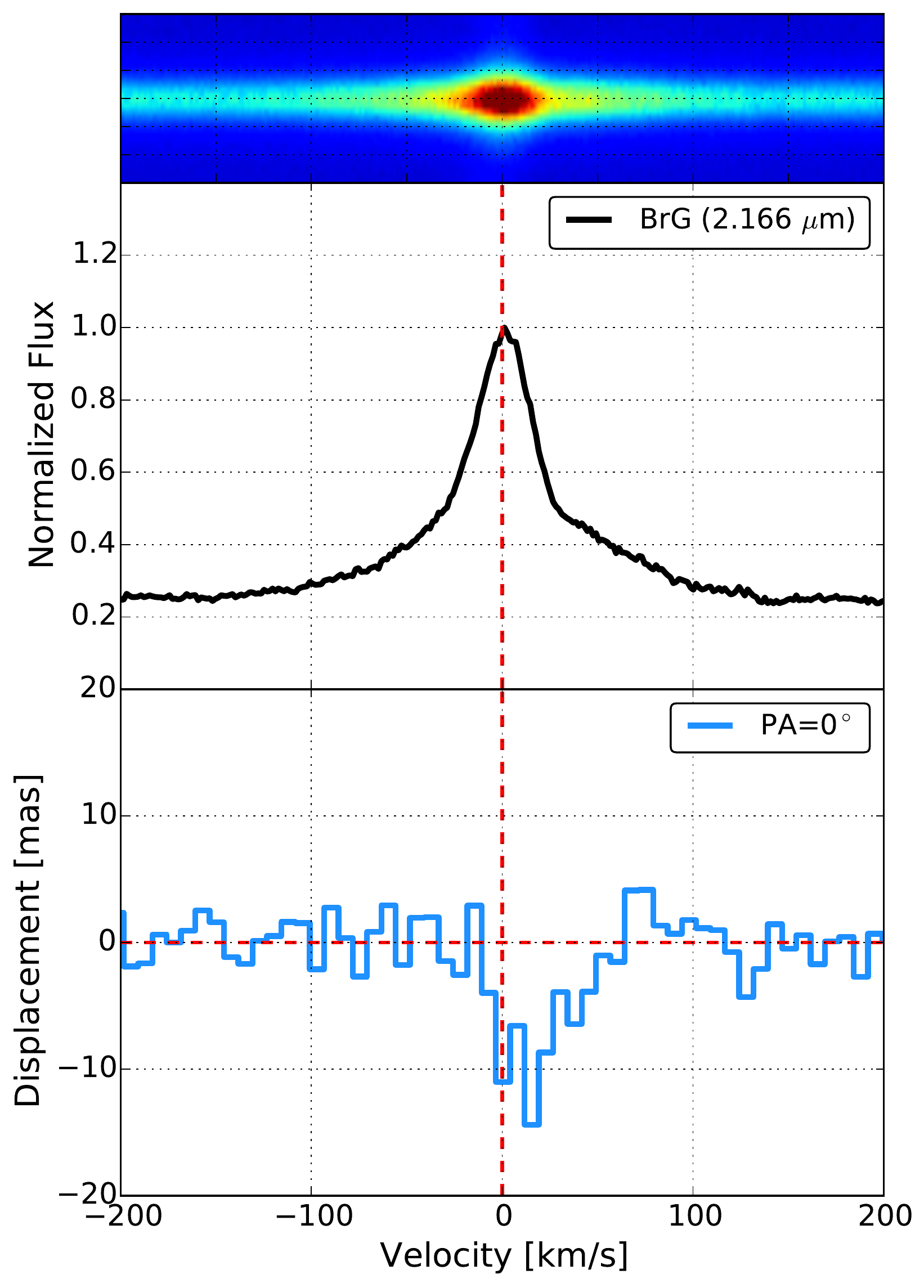} 
\end{minipage}
\begin{minipage}[b]{0.32\textwidth}
\includegraphics[width=5.7 cm, height=6.5 cm]{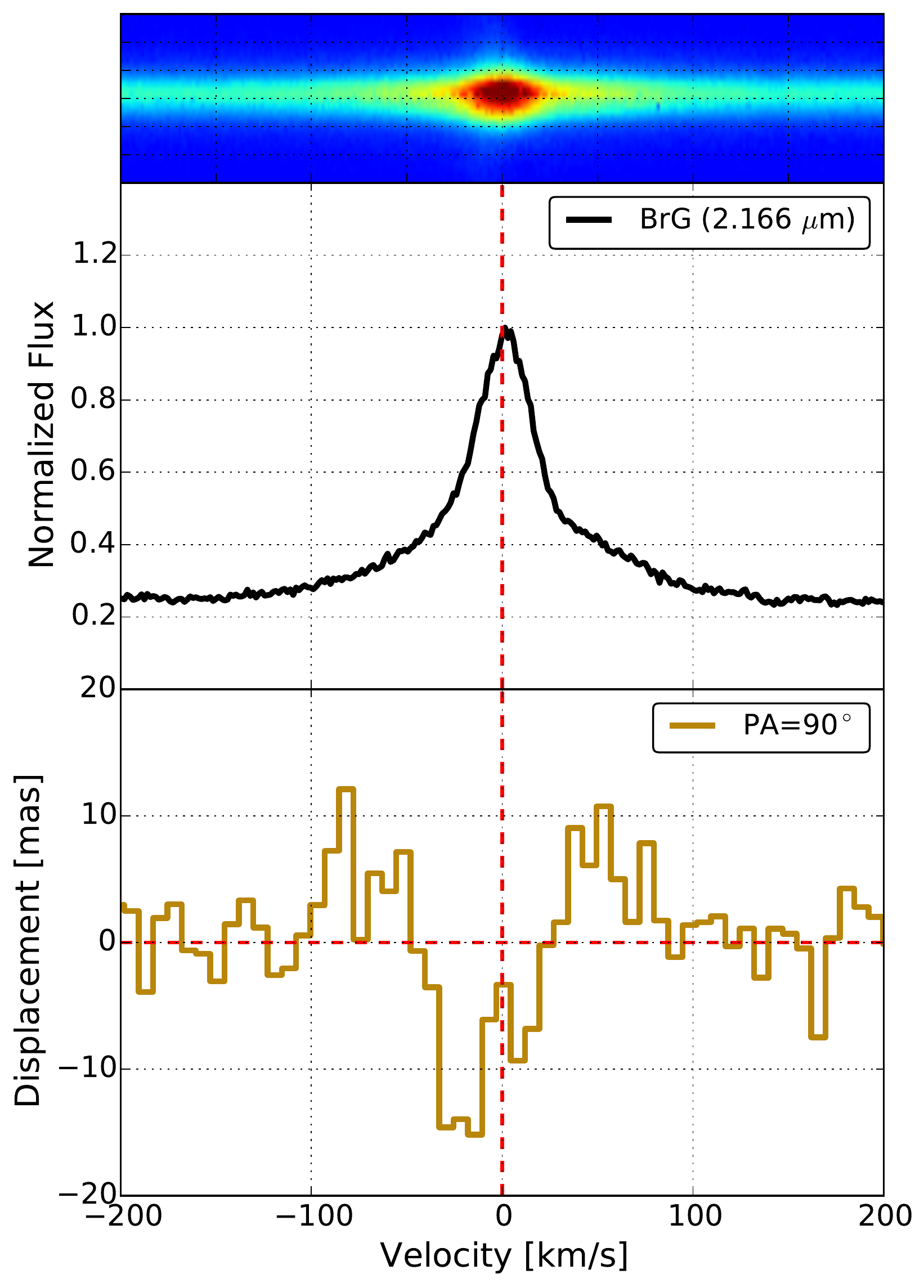} 
\end{minipage}
\begin{minipage}[b]{0.32\textwidth}
\includegraphics[width=5.7 cm, height=6.5 cm]{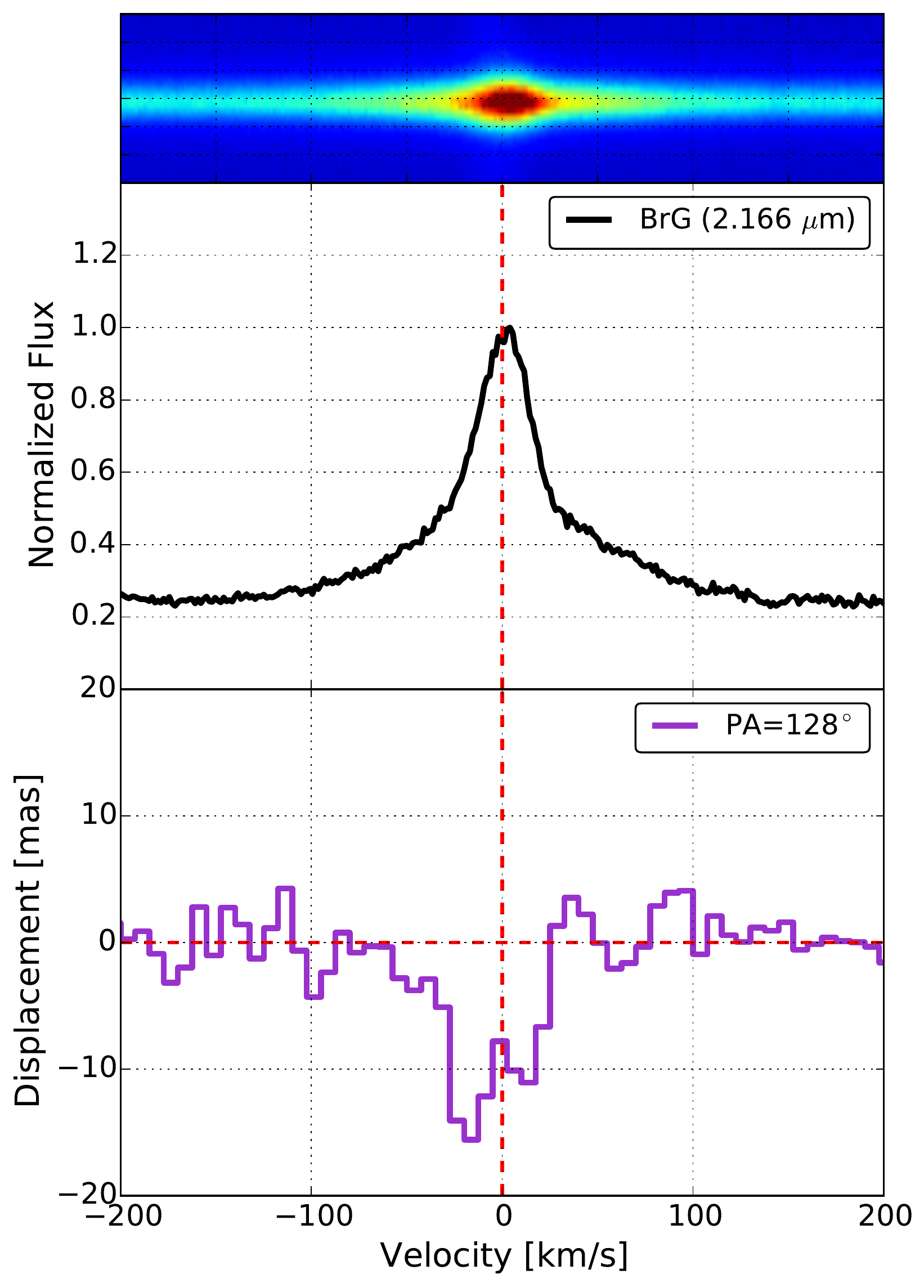} 
\end{minipage}
\end{minipage}

\caption[Spectroastrometric signal of the IRS\,9A Br$\gamma$ and H$_2$ lines]{Spectroastrometric signals of the $H_2$ and $Br\gamma$
  lines of IRS\,9A. The $H_2$ emission line appears to originate from the large-scale structure of IRS\,9A while the $Br\gamma$ line is
  formed at the core of IRS\,9A. The spectroastrometric signal of the line
  displayed is weighted with $w_f$. The mean ratio  $F_{\lambda
    (continuum)}$/$F_{\lambda (line)}$ is of $\sim$0.9 and $\sim$0.5 for the
H$_2$ and the Br$\gamma$ line, respectively. The 2D spectral lines are
also displayed in colors.}
 \label{fig:sa_plots}
\end{figure*}

Figure\,\ref{fig:sa_plots} displays the SA signal of the different
position angles for both $H_2$ and $Br \gamma$ lines. The SA signals,
displayed on the histograms, are averaged over three and five spectral bins for H$_2$ and
Br$\gamma$, respectively. All the signals are corrected for the
continuum effect, using an average  $F_{\lambda
  (continuum)}$/$F_{\lambda (line)}$ of 0.5 and 0.9 for the
Br$\gamma$ and H$_2$, respectively. 

On the one hand,
the $Br \gamma$
emission line appears to be formed at the core of IRS\,9A with maximum offsets from the
continuum of around $\sim$20 mas. On the other hand, in the $H_2$
line, the
SA signal presents maximum offsets of the order of
$\sim$150-300 mas, depending on the position angle of the slit. The sizes of the SA signatures are consistent with the structures observed in our
SAM data. This suggests that the central region of
IRS\,9A has a compact structure that contains ionized material and a
more extended surrounding envelope composed of molecular hydrogen.

The maximum shift in the $H_2$
SA signal at 
PA=0$^{\circ}$ is blue-shifted from the line's systemic velocity, while
the maxima of the shifts in the $H_2$ SA signal at PA=90$^{\circ}$ and
PA=128$^{\circ}$ coincide with the line's systemic velocity. Since the
width of the line profile goes from -10 to 10 km/s, and the
spectral resolution of CRIRES is of 9 km/s, the observed line profile merely
resembles the response function of the spectrograph. There are two main mechanisms that give rise to the
infrared emission of the observed transition, $\nu$=1-0 S(1), in H$_2$: collisional
excitation \citep[e.g.,][]{Smith_1995} and ultra-violet (UV)
resonant-fluorescence \citep[e.g.,][]{Black_1987}. However, to
disentangle which mechanism is responsible for the observed emission
line, a test of
the flux ratio between the H$_2$ line at 2.12$\mu$m and the H$_2$ line
at 2.24$\mu$m ($\nu$=2-1 S(1)) will be required \citep[high ratios between those
lines implies that the origin of the emission is due to shocked gas
and low ratios that the emission is caused by fluorescence; see
e.g.,][]{Wolfire_1991}. 

$Br\gamma$ SA signals at position angles of 90$^{\circ}$ and
128$^{\circ}$ exhibit an inverted double-peaked profile, with one of the peaks
blue-shifted and the other red-shifted. Velocities at the peaks
for both position angles are of about -30 and 30 km/s, respectively. The characteristic SA
signature of a disk-like structure shows two peaks that are reversed
regarding the direction of their spatial displacement \citep[see
e.g.,][]{Pontoppidan_2011_SA, Brown_2013, Blanco_2014}. Therefore, the observed SA
profiles cannot arise from a standard disk-like structure in Keplerian
rotation. On the
contrary, this type of SA signature appears to be formed in more
complex systems with asymmetries. Another intriguing
possibility is that the observed
Br$\gamma$ SA signal could be explained by the presence of a binary system. 
\citet[][]{Binaries_Bailey_SA} presented the characterization of
several pre-main sequence binaries that exhibit similar SA signals to IRS\,9A,
finding double-peak profiles, which had been created by binaries, that variate depending on
the observed PA; with the maximum shift seen at the position angle 
of the binary's orbital plane. Therefore,
binarity of the central source may explain the lack of a clear double-peak
profile at PA=0$^{\circ}$. In this scenario, the observed SA signal
could be explained by the combined effect of two lines, each one
produced by one stellar component, in which one of them exhibits a
narrow profile and the other one a broader line-width \citep[see the case of KK Oph in][]{Binaries_Bailey_SA}. 

\begin{figure*}[htp]
\centering
\begin{minipage}{0.33\textwidth}
\centering
\includegraphics[width=6 cm]{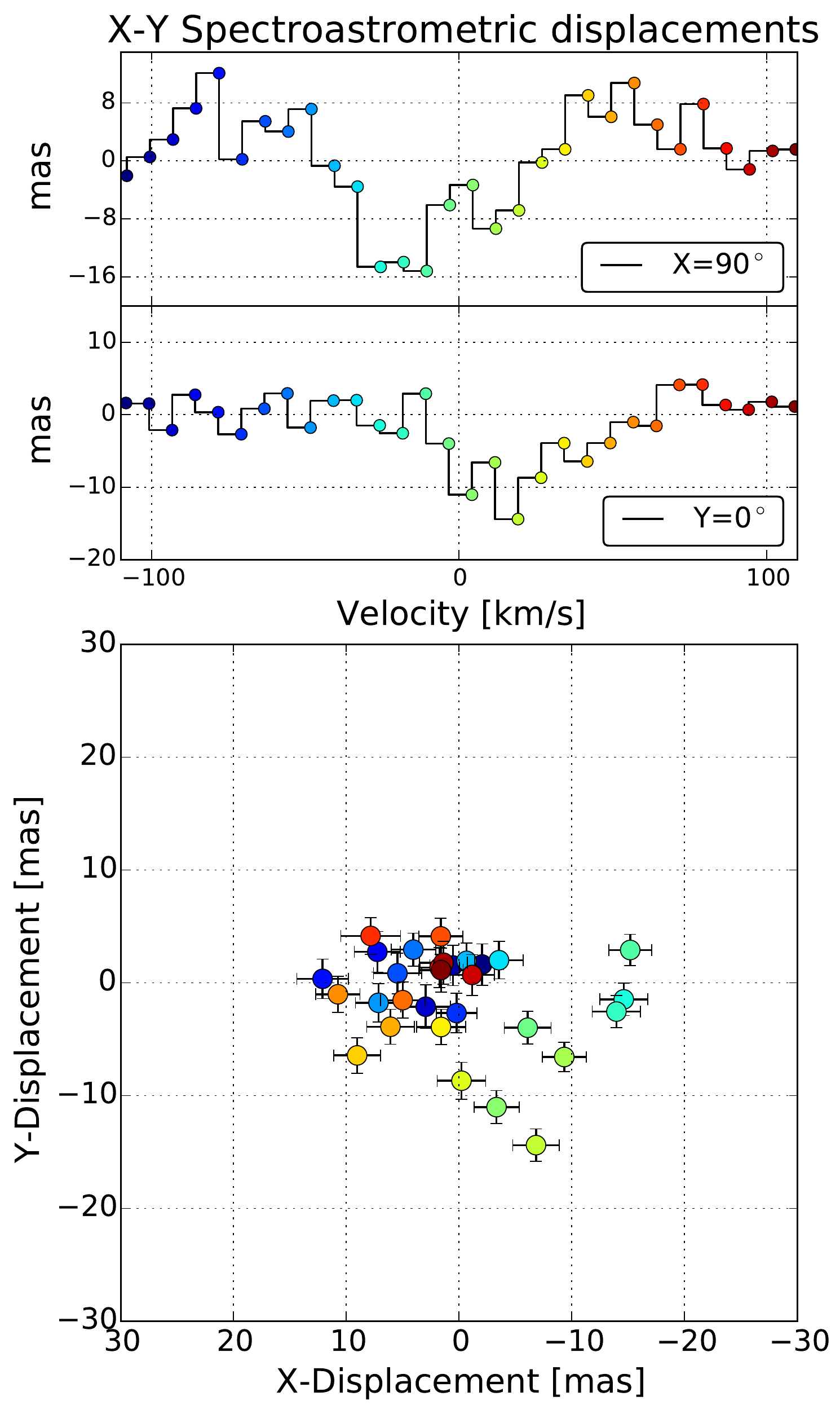}
\end{minipage}\hfill
\begin{minipage}{0.33\textwidth}
\centering
\includegraphics[width=6 cm]{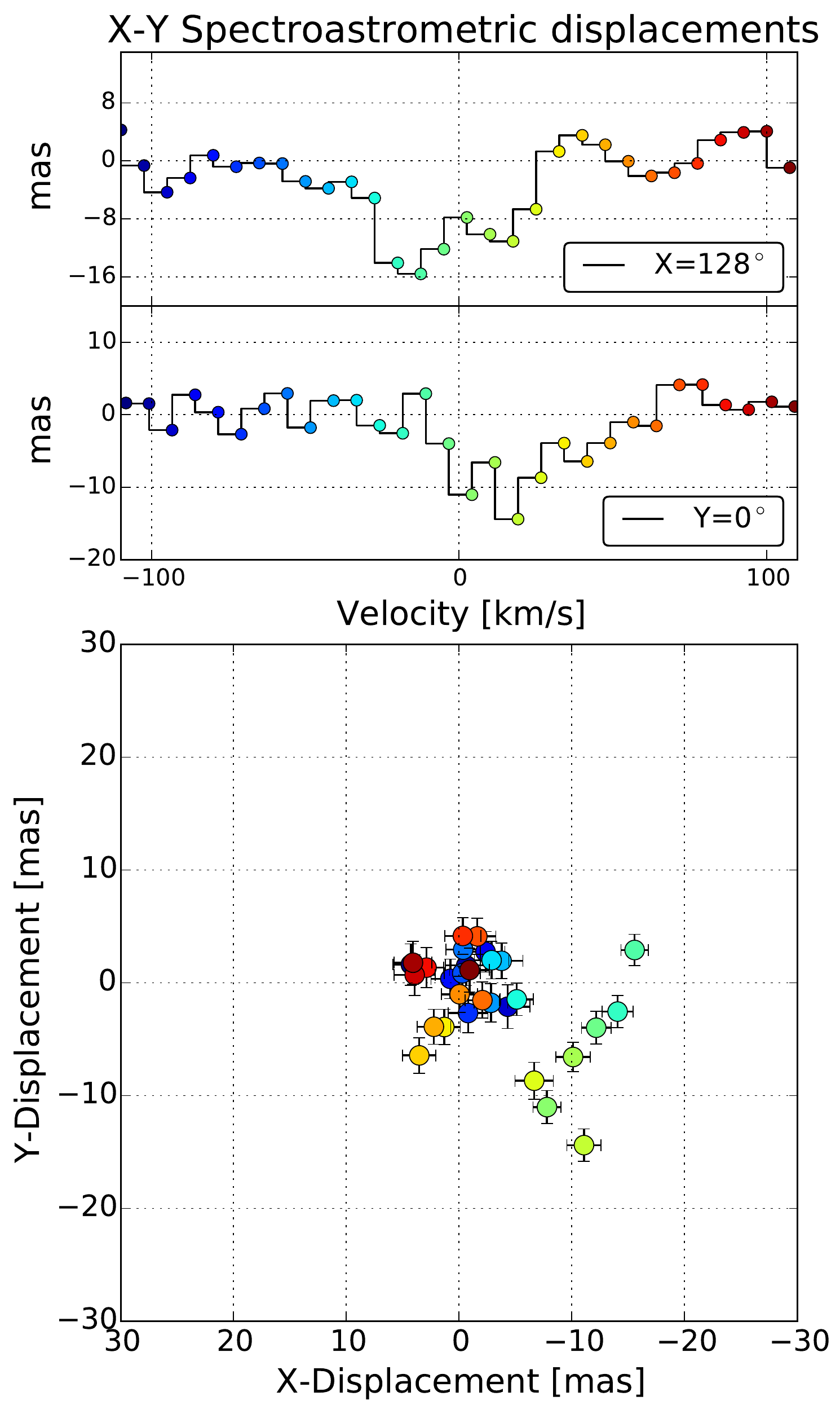} 
\end{minipage}\hfill
\begin{minipage}{.33\textwidth}
\centering
\includegraphics[width=6 cm]{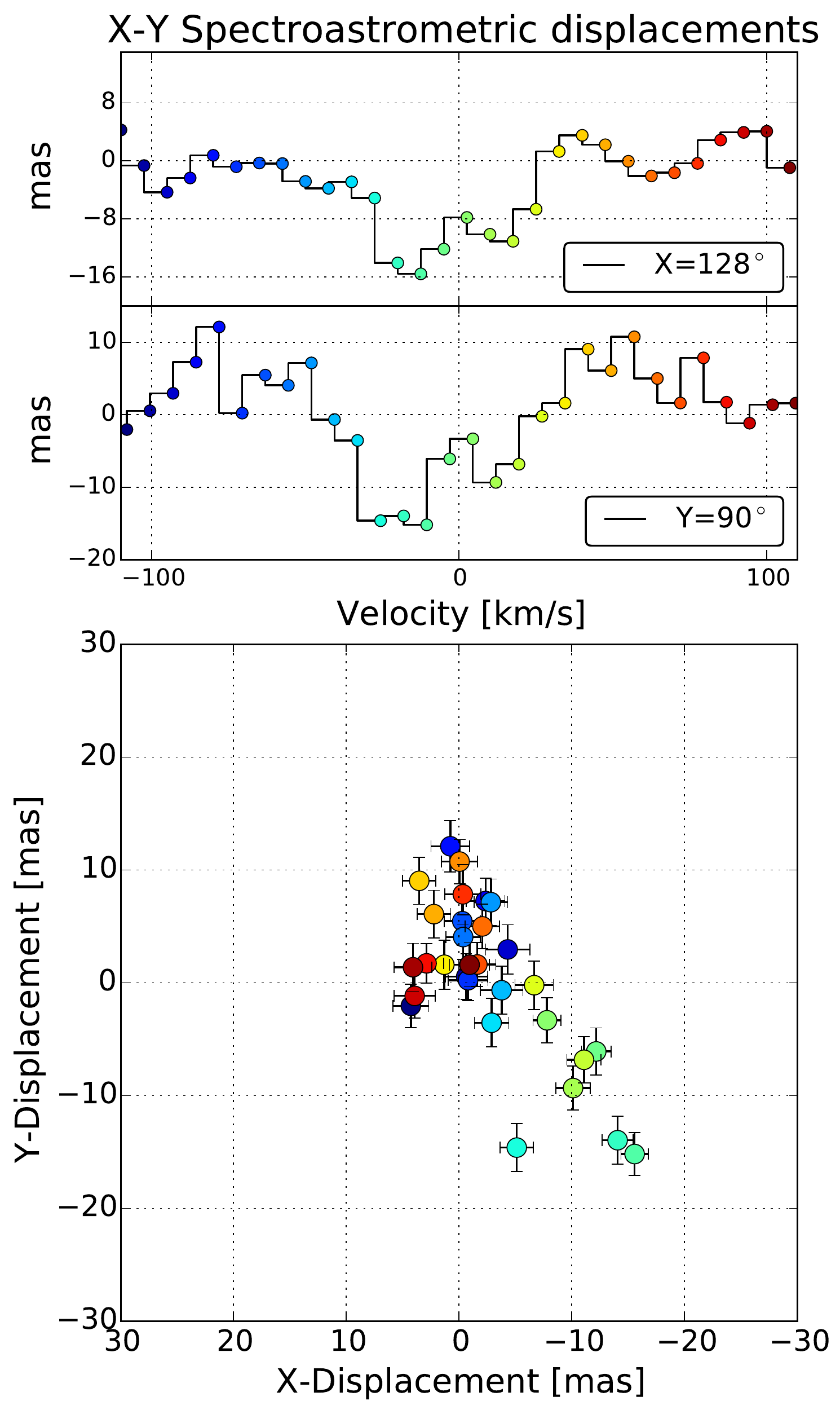}
\end{minipage}\hfill
\caption[2D plots of the Br$\gamma$ spectroastrometric signals]{2D
  plots of the Br$\gamma$ SA signals. Each panel in the columns
  represents the flux centroid displacements at two given position
  angles, as indicated at the top SA signals. Each spectral bin is
  represented by different colors, the uncertainties in the flux
  centroid displacements are also plotted.}
\label{fig:sa_XY}
\end{figure*}

A simple way to determine whether the Br$\gamma$ SA signal is formed of a
distribution of material in a common orbital plane (e.g., disk or
binary) consists in representing two SA signals observed at
different position angles  in an x-y 2D plot. If the resulting trend of the astrometric
offsets in the x-y plot can be approximated by a linear function, then the material (from which the SA signal is formed)
is coplanar, as expected for an orbital plane \citep[see e.g., ][]{Binaries_Bailey_SA, Takami_2003}. Figure\,\ref{fig:sa_XY} displays three
panels with the x-y maps produced using the observed position
angles. The axes plotted in Fig.\,\ref{fig:sa_XY} correspond to the
projection in cartesian coordinates of the sampled position angles. As can be seen, neither the x-y plot created with the SA
signals on position angles 90$^{\circ}$ and 0$^{\circ}$, nor the plot
with position angles  128$^{\circ}$ and 0$^{\circ}$, yields a
linear distribution. The x-y panel formed with PA
  128$^{\circ}$-90$^{\circ}$ shows a correlation among the
  signals. However, we suspect that this is caused by the small
  difference in the displayed position angles. Therefore, we cannot
  confirm a preferential orbital plane at which the SA signals are
  formed. In contrast, the observed SA signals appear to represent the contribution of 
several morphological components of IRS\,9A. Similar findings
  are observed in SA signals of other targets with complex
  morphologies like the Herbig Ae/Be star Z
CMa \citet{Binaries_Bailey_SA}. This object shows at least
two components with different orbital planes, a binary+disk system
with perpendicular outflows, in its 2D spectroastrometric plots. 

\subsection{IRS\,9A radiative transfer model \label{sec:Model}}

\begin{figure}[htp]
\centering
\includegraphics[width=7 cm, height=5cm]{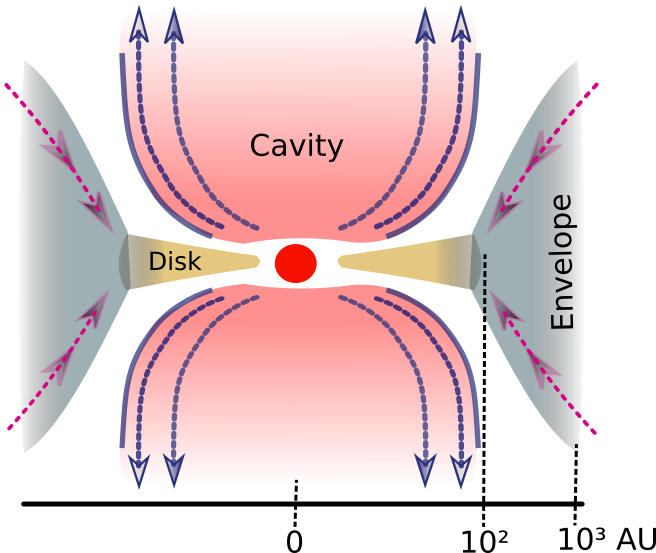} 
\caption[Diagram of the IRS\,9A morphology]{Schematics present
  the adopted model of the IRS\,9A morphology used in our radiative
transfer simulations. The model includes an inner disk surrounded
by an envelope with bipolar cavities. The purple lines in
the cavities indicates the direction of the radiation escaping from
the core, while the pink lines in the envelope represent the
infall of material from the envelope through the disk. The typical
scales of the different components are shown.}
 \label{fig:diagram}
\end{figure}

To determine the physical conditions that best reproduce all the
observational information of IRS\,9A, we linked the observed $V^2$ of our SAM data
with the MIR observations from the literature, and the object's SED through
a multi-wavelength radiative transfer simulation. We considered a physical scenario that includes similar properties
to the Class I YSO described by \citet{Whitney_2003}. The structure of
the model
consists of a compact flared disk around the central
source of radiation that is surrounded by an outer, elongated envelope
with two bipolar cavities that were probably created by
jets and outflows launched at the core of IRS\,9A (see
Fig.\,\ref{fig:diagram}).  A similar morphology was
modeled previously by \citet{Vehoff_2010}. However, we found that
some changes
in the reported parameters should be introduced to fit the new NACO/SAM
data. The 3D density distribution of the disk in cylindrical
coordinates, $\rho_{disk} (R, z, \phi )$, follows a power law
described by the expression:

\begin{equation}
\rho_{disk} (R, z, \phi )=\rho_0^{disk} \left[ \frac{R}{100\,AU}
\right]^{\beta+1}exp\left[-1/2 \left( \frac{z}{h(R)} \right)^2 \right],
\,
\end{equation}
 
where $\rho_0$ is the scale factor of the density, R is the disk
radius, and $\beta$ the scale-height disk exponent. The scale-height
function is given as:
$h(R)=h_0(R/R_0)^{\beta}$. The envelope density
distribution corresponds to an elongated structure of infalling
material, according to \citet{Ulrich_1976}. It has the following form:

\begin{equation}
\label{eq:envelope}
\begin{split}
\rho_{env} (R, z, \phi )=\frac{\dot{M}_{envelope}}{4\pi(GM_*r_o)^{1/2}}
\left[ \frac{r}{r_0}\right]^{-3/2} \left[1+\frac{\mu}{\mu_0}\right]^{-1/2}\left[ \frac{\mu}{\mu_0}+\frac{2\mu_0^2r_0}{r}\right]^{-1} ,
\,
\end{split}
\end{equation}  

where r is the radius of the envelope, r$_0$ is the normalization radius
of the envelope, and $\dot{M}$ is the rate of mass infall. The factors $\mu_0$ and
$\mu$ are related by the equation for the streamline:
$\mu_0^3+\mu_0(r/r_0-1)-\mu(r/r_0)=0$. The density distribution in the envelope cavity was considered to be
uniform. It is important to emphasize that our simulation does not
consider binarity and/or irregularities in the disk at the core of
IRS\,9A; such detail lies beyond the scope of this work. 

To perform the radiative transfer
simulation, we used the \textit{Hyperion} software \citep{Hyperion_2011}. This code carries out 3D dust-continuum
radiative transfer simulations while creating SEDs and images
at the required wavelengths. Our \textit{Hyperion} simulations used the
density distributions of the previously mentioned structures to determine their
temperature and flux maps. The code uses
a modified random walk
(MRW) approximation to propagate the photons in the thickest regions
of the simulations. Our model assumes a modified MRN grain-size distribution
\citep{Mathis_1977} as determined by \citet{Kim_1994}. The chemical
composition of the dust includes astronomical silicates, graphite, and carbon. The simulated models also include the scattering of the dust with a full
numerical approximation to the Stokes parameters in the
ray-tracing process. The gas-to-dust ratio of 100:1, a foreground
extinction of Av=4.5 (the extinction law used is described in
\citet{Robitaille_2007}), and a distance of 7 kpc
\citep{Nurnberger_2003} were used
for all simulations. 

From the radiative transfer simulations, we obtained synthetic images
of IRS\,9A for the $K_s$ and $L'$ bands, observed with NACO/SAM (see Sec.\,\ref{sec:NACO_SAM}), 
for the $N-$band (11.7 $\mu$m), observed with T-ReCS, and for the MIDI/VLTI (8-13
$\mu$m) data, described in \citet{Vehoff_2010}. Subsequently, the $V^2$ were
extracted from those images, at the sampled spatial frequencies by the
observations, using proprietary IDL/Python routines. We also obtained
synthetic SEDs in the range of 1-500 $\mu$m. For all images, the
total flux was normalized according to the angular size of the primary beam.

\begin{table*}[ht]
  \caption{Radiative transfer models of IRS\,9A}
\label{tab:models}
\centering
\begin{tabular}{l c c c}
\hline\hline
Parameter & Best SED-only fit $^{\mathrm{a}}$ & Vehoff+2010 & Best
$V^2$ model \\
 & (No. 3006825)$^{\mathrm{b}}$ & (No. 3012790)$^{\mathrm{b}}$ & (this
                                                                 work)
  \\ \cline{4-2}
\hline
Total luminosity (L$_{\odot}$) &  1.0x10$^5$ & 9.2x10$^4$ & 8.0x10$^4$\\
Stellar mass (M$_{\odot}$) & 27.2 & 25.4 & 30.0 \\
Stellar temperature (K) & 39000 & 38500& 38000\\
Disk mass$^{\mathrm{c}}$  (M$_{\odot}$) & 1.8x10$^{-1}$ &
5.0x10$^{-1}$ & 5.0x10$^{-1}$\\
Disk outer radius (AU) & 91.0 & 94.0 & 80.0 \\
Disk inner radius (AU) & 28.0 & 25.0 & 10.0 \\
Disk flaring power -$\beta$- & 1.1 & 1.2 & 1.2\\
Disk scale height at 100 AU -$h$-& 8.9 & 8.8 & 8.0\\
Envelope outer radius (AU) & 1.0x10$^5$ & 1.0x10$^5$ & 7x10$^3$\\
Envelope cavity angle (deg.)$^{\mathrm{d}}$ & 31.0 & 29.0 & 30.0\\
Bipolar cavity density (g/cm$^3$) & 7.7x10$^{-21}$ & 9.9x10$^{-21}$ & 1.0x10$^{-20}$\\
Inclination -$i$- (deg)$^{\mathrm{e}}$ & 87.0 & 85.0 & 60.0 \\
Position Angle -$\phi$- (deg)$^{\mathrm{f}}$ & 105.0$^{\mathrm{g}}$ & 105.0 & 120.0 \\
\hline
\end{tabular}
\begin{list} {}{} \itemsep1pt \parskip0pt \parsep0pt \footnotesize
\item[$^{\mathrm{a}}$] This is the best-fit model to the SED obtained
  with Robitaille's online fitting tool.
\item[$^{\mathrm{b}}$] Number of the model from Robitaille's database.
\item[$^{\mathrm{c}}$] Total mass (dust+gas).
\item[$^{\mathrm{d}}$] Half-opening angle of a conical cavity in the envelope.
\item[$^{\mathrm{e}}$] Inclination angle of the disk-envelope from the line of sight:
  0$^{\circ}$ means that the disk and envelope are seeing face-on,
  90$^{\circ}$  means that the disk and envelope are seeing edge-on.
\item[$^{\mathrm{f}}$] Position angle of the disk mid-plane projected
  in the plane of the sky, measured East of North.
\item[$^{\mathrm{g}}$] Assumed to be the same as \citet{Vehoff_2010}.
\end{list}
\end{table*}

\subsubsection{The parameter space \label{sec:parameter_space}}

\begin{figure}[htp]
\begin{minipage}{0.5\textwidth}
\centering
\includegraphics[width=6 cm]{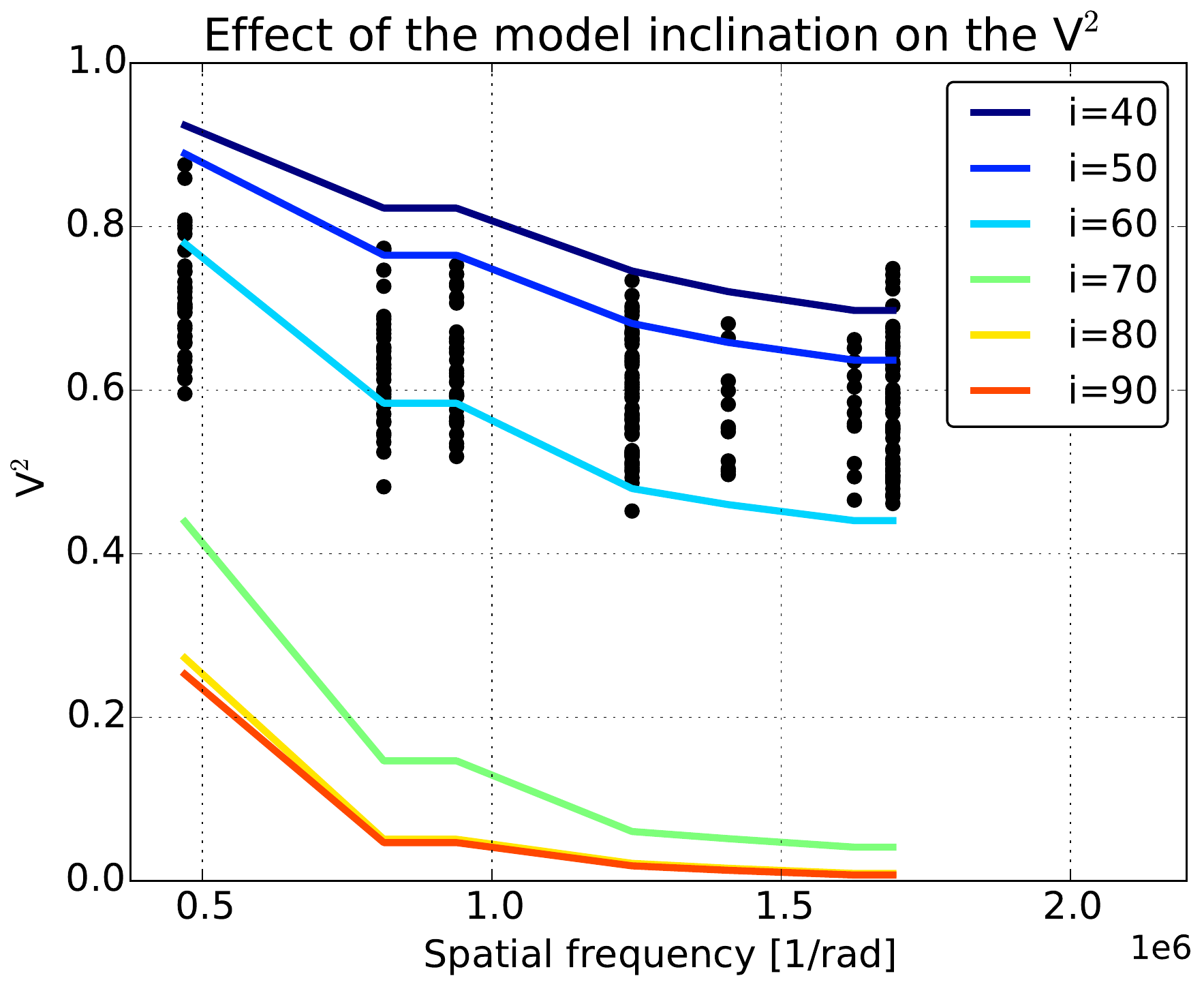}
\end{minipage}\hfill
\begin{minipage}{0.5\textwidth}
\centering
\includegraphics[width=6.1 cm]{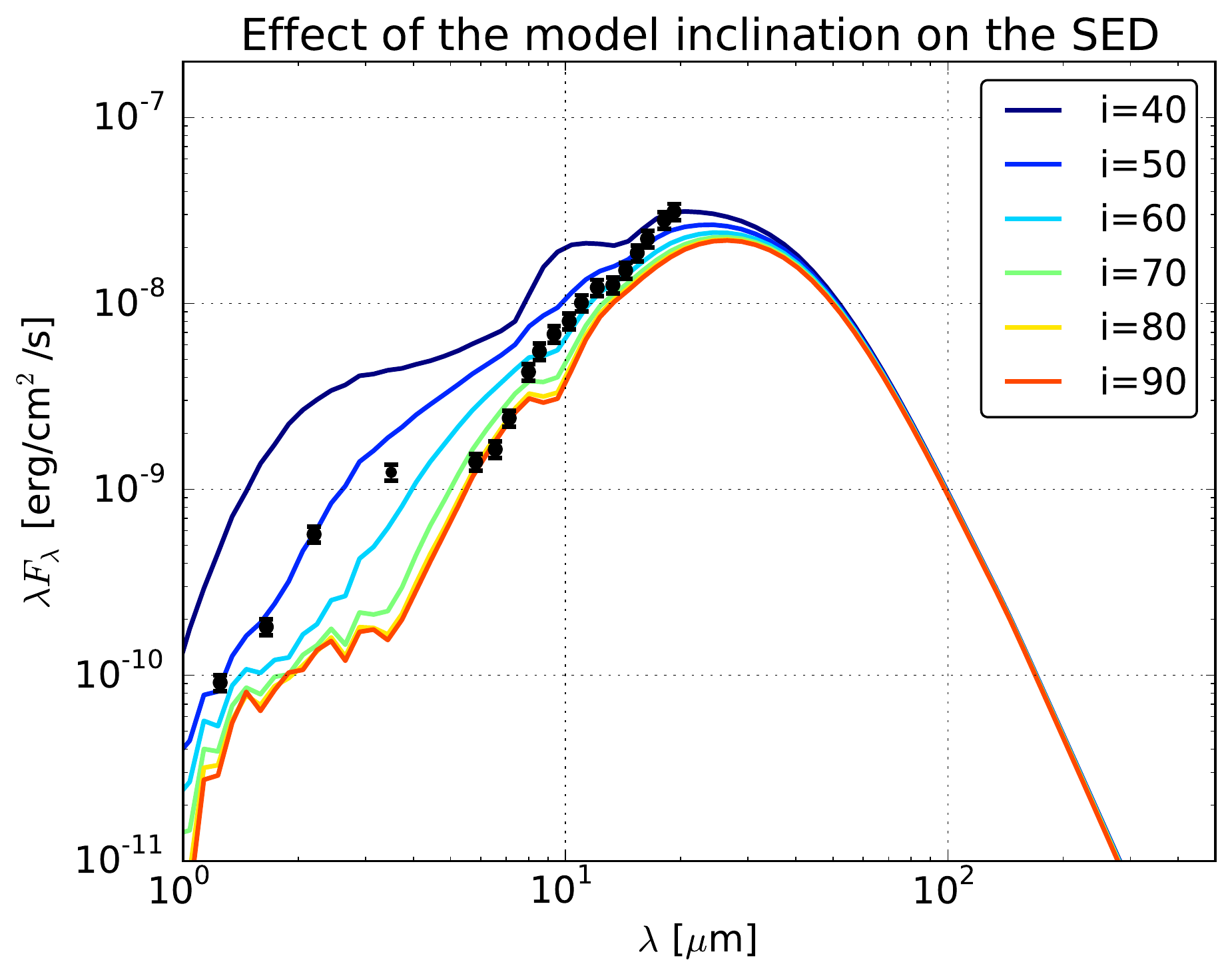} 
\end{minipage}\hfill
\caption[Radiative transfer models of IRS9A at different
inclinations]{Effect of changing in the inclination angle, $i$, of the
  model on the simulated
  $V^2$ and SED. On the \textit{upper} panel, the effect of modifying $i$ on the
  NACO/SAM $L'-$filter is observed. On the \textit{lower} panel, the
  effect of changing $i$ over the SED is displayed. The different
  colors correspond to different values of $i$, while the black dots
  represent the data in both panels. In this example, the rest
  of the model parameters remained fixed to the values of the best-fit model
  presented in this work}
 \label{fig:irs9a_incl}
\end{figure}

To constrain the parameters of our simulation, as a
  starting point, we used the disk-envelope Vehoff's
model. However, we observed that the new SAM data were not fit by this
model. Therefore, some adjustments were necessary to reproduce the
observed visibilities.  Our intent was not to revisit the full
parameter space because the grid of SED models computed by
\citet{Robitaille_2006} has sampled it quite sparsely for
MYSOs. Beginning with Vehoff's model, we tuned each of the parameters
in turn by hand, starting with the inclination and inspecting the fits
to visibilities and SED. The tuned parameters are as follows:

\textit{a) Orientation ($i$, $\phi$):} Inclination angle, $i$, of the disk mid-plane
from the line of sight was identified as a critical parameter for our modeling. It was noticed that changes in this parameter
have a strong impact on the modeled visibilities of the SAM data, besides the
changes in the SED and MIR wavelengths. Values of $i$ close to
90$^{\circ}$ produced systematically low visibilities of the sampled SAM spatial
frequencies. On the other hand, values of $i\sim$60$^{\circ}$ produced
simulated visibilities closer to the expected SAM values.
Figure\,\ref{fig:irs9a_incl} displays the effect of inclination on the simulated V$^2$ for the $L'-$filter and
SED for different values of $i$ between 40$^{\circ}$ and 90$^{\circ}$
in steps of 10$^{\circ}$. From these results, we found that an
average value of $i$=60$^{\circ}$ best-fit the different data sets. To obtain the position angle, $\phi$, of the
mid-plane of the disk in the plane of the sky, we rotate our projected
model in intervals of 10$^{\circ}$ from $\phi$=0$^{\circ}$ to
$\phi$=360$^{\circ}$. Since the lack of phase information in the MIDI
data and that the NACO/SAM closure phases were not sensitive to the
projected position
angle (because IRS\,9A was only partially resolved; see Fig.\,\ref{fig:v2_calib}),
we adopted the value of $\phi$ that minimizes the residuals of V$^2$
and CPs of the T-ReCS data. In this case $\phi$=120$^{\circ}$ (see
Figs.\,\ref{fig:CPs_TRECS} and \ref{fig:RGB}).

\textit{b) The central source ($T_{eff}$, $L_*$, $M_*$):} As
established in \citet{Vehoff_2010}, the model
assumes the typical stellar parameters of a main-sequence O-star \citep[see e.g.,][]{Martins_2005}. The
effective temperature was thus fixed at
$T_{eff}=$38000 K and
the mass to M$=$30M$_{\odot}$. Only variations of the luminosity were
performed in steps of 1x10$^4L_{\odot}$ from $L_*=$1.0x10$^5L_{\odot}$
to 8x10$^4L_{\odot}$ to have a reasonable fitting to the SED.

\textit{c) The disk ($R_{in}$, $R_{out}$, $m_{disk}$, $\beta$, $h_0$):} The inner radius of the disk was obtained
from variations of the value given by \citet{Vehoff_2010}. The covered
values went from 10 to 25 mas in increments of 5 mas. The value that
best reproduce the observed $V^2$ is $R_{in}$=10 mas. This value is
in agreement with the sublimation radius of the dust (for a sublimation
temperature $T_{sub}=$1500 K) according to
\citet{Monnier_2002}:

\begin{equation}
\label{eq:irs9a_subr}
R_{sub, disk} =1.1\sqrt{Q_R}\left[\frac{L*}{1000 L_{\odot}
  }\right]^{1/2}\left[\frac{T_{sub}}{1500K} \right]^{-2}~AU \,
\,
\end{equation}

Considering a ratio of the dust absorption efficiencies $Q_R$=1, the inner
radius was thus estimated as: $R_{in}$=$R_{sub, disk}$=10 AU. On the
other hand, the initial value of the outer radius was fixed to
$R_{out}$=100 AU, since this was the value derived from the geometrical
model (R$\sim$14 mas $\sim$ 98 AU) described in
section\,\ref{sec:NACO_SAM}. Variations of $R_{out}$ were performed in
steps of 10 AU from 40 to 100 AU. We observed that changes in this
range do not have a strong impact on the simulated SED, and that they only
introduced slight changes in the simulated $V^2$ within the scatter of the
observed $V^2$. Therefore, we adopted an average value of $R_{out}$=80 AU. For the mass of the disk and the
flaring exponent, $\beta$, we used the same values as \citet{Vehoff_2010}, thus
$m_{disk}$=5x10$^{-3}$M$_{\odot}$ and $\beta$=1.2, respectively. To
determine the scale height, $h_0$, at 100 AU, we performed several tests
with values of $h_0$ between 3-10 AU, with steps of 1 AU, finding that
the value that best reproduces the visibilities is $h_0$ =8 AU, which
is very
close to the value previously reported by \citet{Vehoff_2010}.

\begin{figure*}[ht]
\centering
\includegraphics[width=16 cm, height=12 cm]{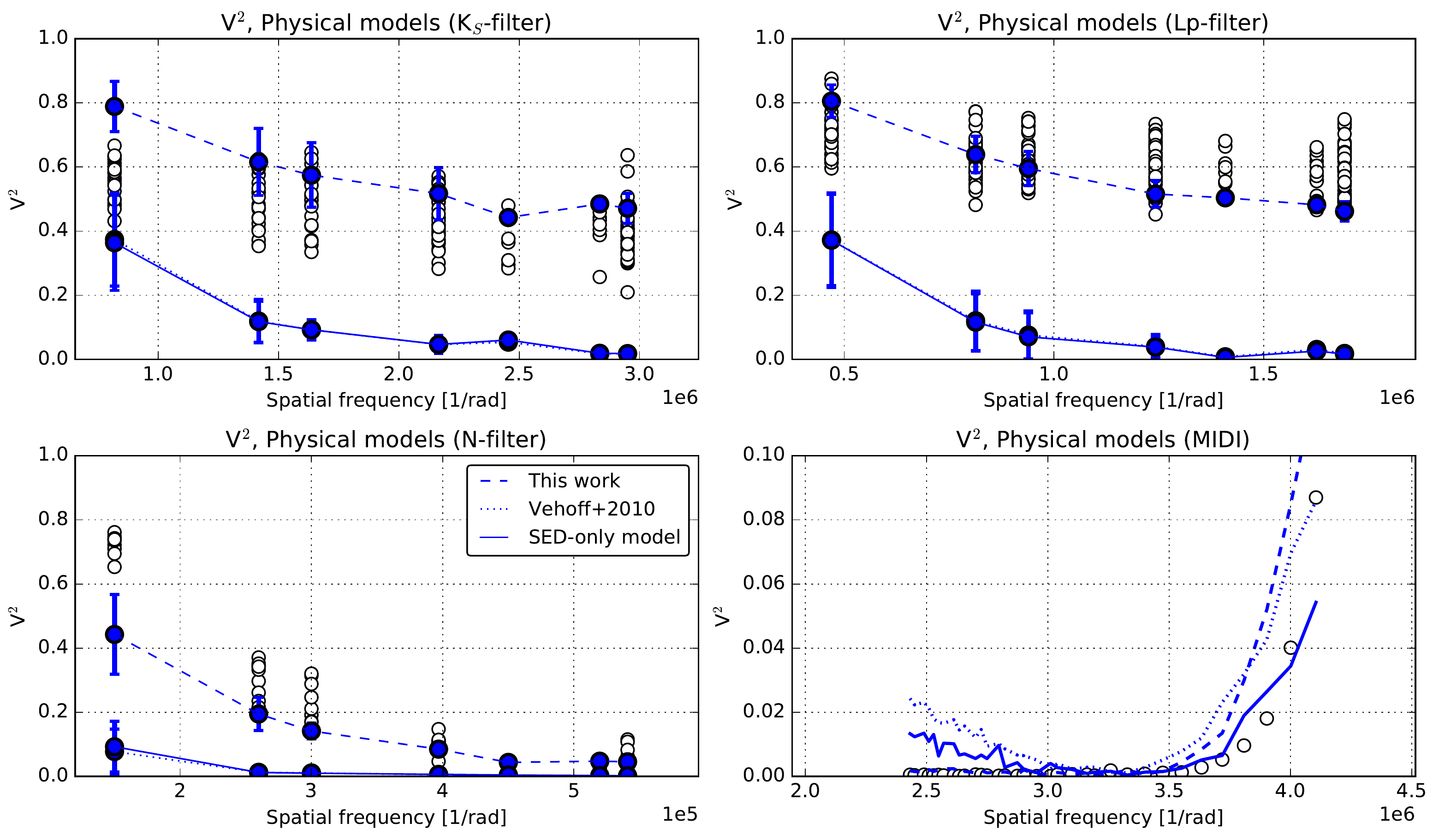} 
\caption[Radiative transfer models of IRS9A (V$^2$) ]{Models of IRS\,9A, giving predicted $V^2$ at NIR and MIR wavelengths. The
  model signals are displayed with varying linetypes (see key). The
  data are displayed with open black circles. At each spatial frequency for the
  $K_s$, $L'$, and $N$ data sets, the mean value of the model
  visibility is shown with a filled blue circle, and with a vertical
  bar that indicates the variation as a function of azimuth angle.}
 \label{fig:irs9a_models}
\end{figure*}

\textit{d) Envelope and cavities ($\phi$, $\rho_{cav}$, $R_{out,cav}$):}
The parameters of the envelope and cavities were adopted from
\citet{Vehoff_2010}. The value of the opening angle of the cavity
($\phi$) was fixed to
30$^{\circ}$, and the density ($\rho_{cav}$) to 1x10$^{-20}$
g/cm$^{3}$. For the envelope, we adopted a normalization radius of 100
AU and an outer radius of 7x10$^3$AU (1"). The scale of the outer
radius was inferred from the extension observed at MIR wavelengths,
which exhibit an angular scale of around 1''. Tests with larger
$R_{out,cav}$, like the value reported by \citet{Vehoff_2010}, produced
visibilities systematically  below the observed levels, and were therefore
discarded.

Figure\,\ref{fig:irs9a_models} presents a comparison of the V$^2$
fitting of (i) our best-fit model
with (ii) the model of the MIR morphology of IRS\,9A presented by \citet{Vehoff_2010}, and with (iii) the best-fit model of the IRS\,9A
SED that was obtained with Robitaille's online fitting
tool\footnote{http://caravan.astro.wisc.edu/protostars/}
\citep{Robitaille_2007}. Vehoff's model was also obtained from Robitaille's tool, but it does not correspond to the best-fit
model delivered by the online database for the used data, although it
does provide a
reasonable fit to the MIR visibilities. We included both models in Fig.\,\ref{fig:irs9a_models} for
completeness. Remarkably, the model of
\citet{Vehoff_2010}, analyzed here, has a maximum angular size at zero
spacing of $\sim$2''. Caution is recommended when 
making comparisons to Fig.~6 of \citet{Vehoff_2010}, since these authors
limited the maximum size to
0.85'', which lead to bias in the simulated T-ReCS visibilities of the shortest
baselines. Table\,\ref{tab:models} displays the parameters of each
model. Figure\,\ref{fig:SED} displays the SED of IRS\,9A for
each of the models plotted over the observational data. The NIR
photometry was taken from \citet{Nurnberger_2003} and the MIR one from
\citet{Vehoff_2010}. Fig\,\ref{fig:RGB} displays an RGB image,
composed from our models at 2.2, 3.8, and 11.7 $\mu$m.

\begin{table*}[ht]
  \caption{Comparison of the root mean square error (RMSError) between the different
    models of IRS\,9A}
\label{tab:RMSR}
\centering
\begin{tabular}{l c c c}
\hline\hline
Data set & Best SED-only fit$^{\mathrm{a}}$ & Vehoff+2010 & Best
$V^2$ model \\
 & (No. 3006825) & (No. 3012790) & (this
                                                                 work)
  \\ \cline{4-2}
\hline
NACO/SAM $K_s-$filter & 6.3 & 6.3 & 3.1\\
NACO/SAM $L'-$filter & 25.2 & 25.4 & 4.9\\
T-ReCS data & 20.4 & 20.0 & 9.4 \\
MIDI data & 8.4 & 4.4 & 3.5\\
SED & 2.29 & 2.98 & 3.4 \\ 
\hline
\end{tabular}
\begin{list} {}{} \itemsep1pt \parskip0pt \parsep0pt \footnotesize
\item[$^{\mathrm{a}}$] This is the best-fit model to the SED, obtained
  with Robitaille's online fitting tool.
\end{list}
\end{table*}

\begin{figure}[htp]
\centering
\includegraphics[width=\columnwidth]{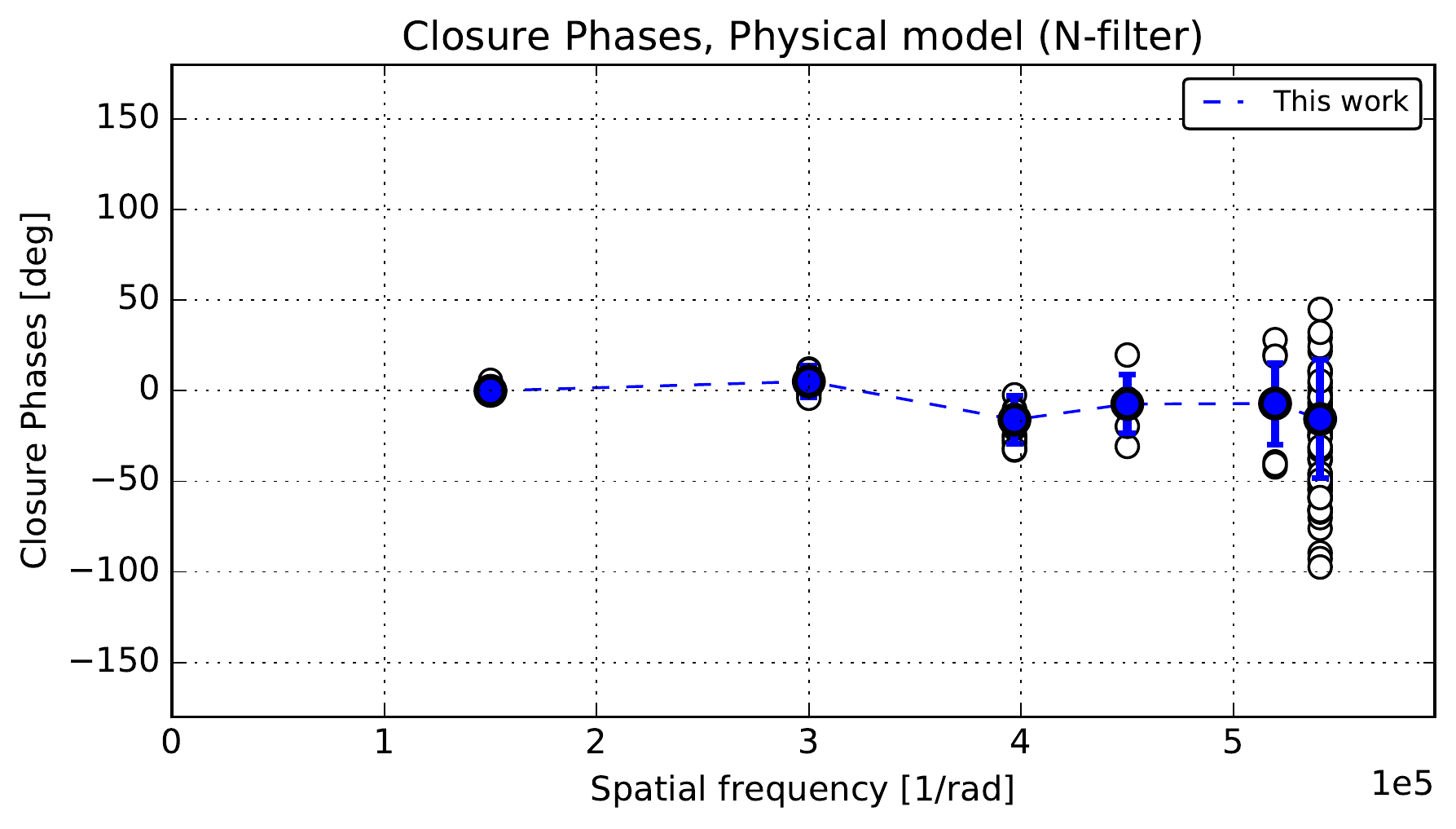} 
\caption[Best-fit model to the T-ReCS CPs]{ Best-fit of our radiative
  transfer model to the T-ReCS closure phases. The
  data are displayed with open black circles and the mean value of the model
  closure phase is shown with a filled blue circle and with a vertical
  bar that indicates the variation as a function of azimuth angle.}
 \label{fig:CPs_TRECS}
\end{figure}

\section{Discussion \label{sec:Discussion}}

As shown in Fig.\,\ref{fig:irs9a_models}, the
model presented by \citet{Vehoff_2010}, which is based on the MIR data, does not
reproduce the observed $V^2$ signals extracted from the new NIR SAM observations. 
In fact, that model is shifted systematically
towards lower $V^2$ values at the three $K_s$, $L'$, and $N$
bands. Similar $V^2$ misfits have been observed for the model
obtained with Robitaille's online tool, which was based only on the SED data. To perform a more quantitative comparison between the different
models, we computed the root mean squared error
(RMSError)\footnote{$RMSError=\sqrt(\frac{\Sigma_i(((d_i-m_i)/\sigma_i)^2)}{n})$,
  where $d_i$ corresponds to the sampled data point $i$; $m_i$ is the
  corresponding model point; $\sigma_i$ is the 1-$\sigma$ uncertainty
  of individual data points; $n$ is the total number of the sampled points.}
for each $V^2$ data set and SED. Table\,\ref{tab:RMSR} displays the
obtained RMSError for the different models and data sets. We note how the
model presented in this work exhibits considerably lower residuals than the other
two models for the NACO/SAM and T-ReCS data. However, the residuals of
our model are similar to the ones presented by the other two models
for the MIDI data.

\begin{figure}[htp]
\centering
\includegraphics[width=\columnwidth]{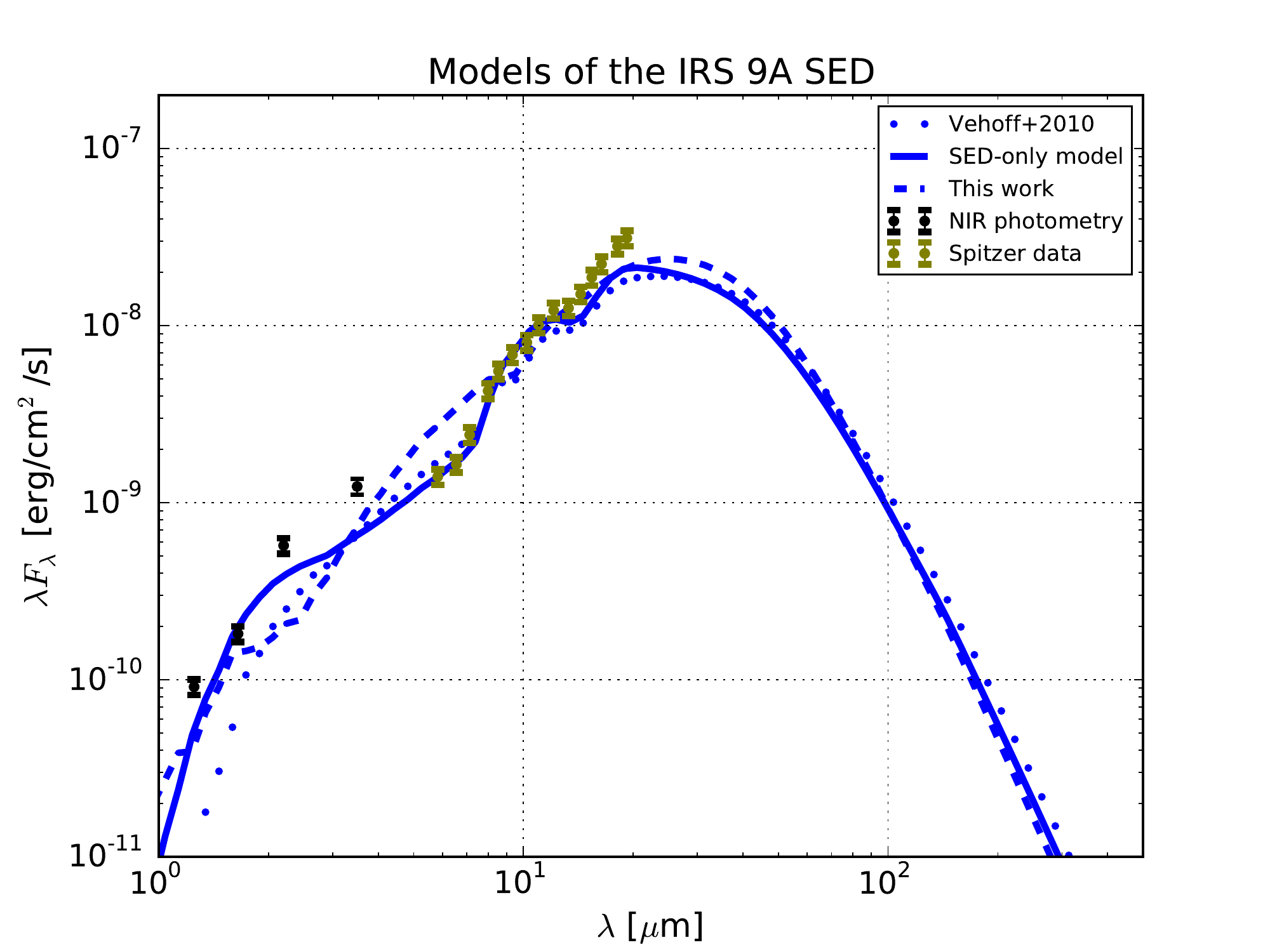} 
\caption[Radiative transfer models of IRS9A (SED) ]{Comparison between
  the various different so-called best-fit (see text) models 
 to the IRS\,9A SED. The photometry data are displayed in black and green. The
  different models are shown in blue (see the caption inside the plot). The
SED models are displayed, assuming an aperture of 3'' at a distance of
7.0 kpc and A$_v$=4.5. }
 \label{fig:SED}
\end{figure}

\begin{figure}[htp]
\centering
\includegraphics[width=8 cm]{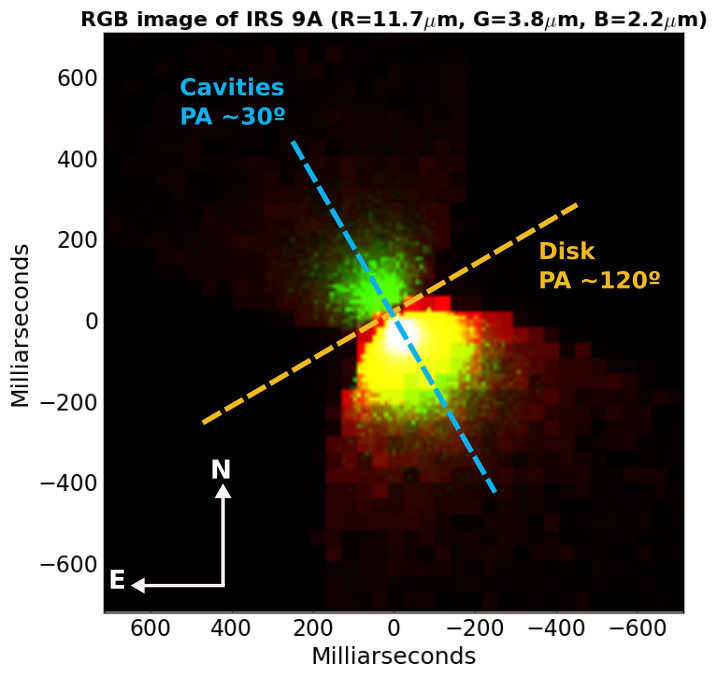} 
\caption[Simulated RGB image of the IRS\,9A morphology]{RGB image
  composed with our model. Note how the emission arises from the
  cavities that dominate the IR. The red, green, and blue filters correspond
to the images of the models at 11.7 $\mu$m, 3.8 $\mu$m, and 2.2
$\mu$m. The position angles (measured East to the North) of the disk and cavities in the plane of
the sky are shown.}
 \label{fig:RGB}
\end{figure}

From Fig.\,\ref{fig:SED} it is appreciated that, despite this poor performance when confronted with the interferometric data, 
good fits to the SED are exhibited by both Robitaille and Vehoff
models. In fact, our model presents the
highest residuals. Nevertheless, 
this result is not unexpected since SED fitting alone often offers highly 
degenerate outcomes: strong constraints are only provided when SEDs are used
in concert with spatial information that identifies the origin of the emission. 
For example, even single envelope models are sufficient to
reproduce the observed SED, although none reproduce, simultaneously, the
observed $V^2$. Furthermore, the apertures used to extract the SED
measurements are considerably larger (3'') than the angular size of
IRS\,9A. Therefore, some measurements may be contaminated by flux from
nearby sources in the field \citep[see, e.g., Fig\,2 in ][in which additional sources are observed
around IRS\,9A in a radius of $\sim$3'']{Nurnberger_2008}. This underlines the importance of obtaining SED data at adequate
angular resolution, as well as multi-wavelength interferometric observations, combined with SED modeling, to construct a
reliable physical framework of the MYSOs morphology. 

In contrast to previous works, we have produced a model of
IRS\,9A that reproduces all observable data, including $V^2$ signals at 
NIR and MIR wavelengths (although some deviations are present at the shortest 
baselines of the $K_s$ and $L'$ filters and in the largest baselines of the MIDI data). 
Our model reproduces the longest baseline visibilities from the T-ReCS data, but underestimates
V$^2$ at the shortest T-ReCS baselines. 
We emphasise that, in contrast to the other filters,
the N-band T-ReCS data sample the largest scale of IRS\,9A. Therefore, we infer that our model faithfully reproduces IRS\,9A morphology up
to scales $\le$ 1'' (7000 AU), but significantly differs from the
V$^2$ data that correspond to angular scales between 1''-1.6'' (7000-11000 AU). In
fact, from the reconstructed T-ReCS/SAM image presented in
\citet{Vehoff_2010}, it is noticeable that the morphology of IRS\,9A
at larger scales is irregular and more complex than our
modeling. 

\begin{figure}[htp]
\centering
\includegraphics[width= 8cm]{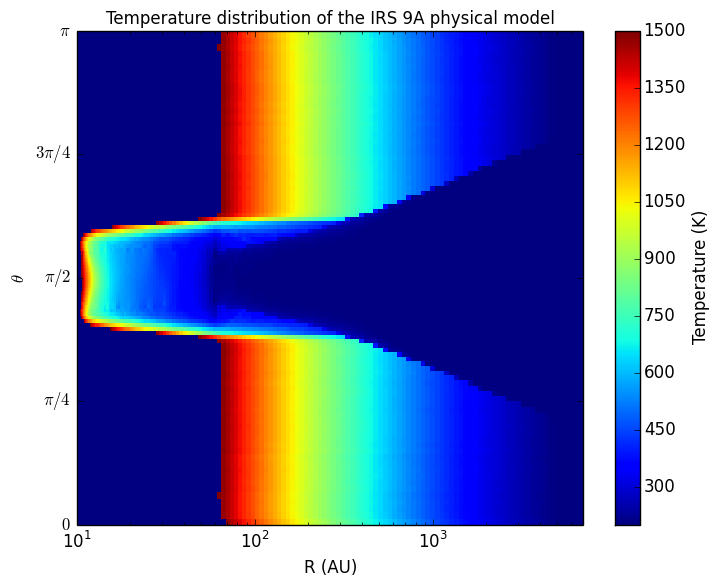} 
\caption{Radial cut of the temperature distribution of the dust around
  IRS\,9A. The picture displays the temperature profile of the disk
  and the envelope obtained with our best radiative transfer model. The axis is given in
spherical coordinates (R, $\theta$) and the colors represent
different temperature values of the dust. Also, confusion with other
(extended or stellar) sources in the crowded IRS\,9 region may be an
important source of bias in the SED at certain wavelengths.}
 \label{fig:temp}
\end{figure}

Figure\,\ref{fig:temp} displays a radial slice through the temperature
distribution of our best model. It is observed that the disk
atmosphere has a temperature close to the dust sublimation limit (T$\sim$1500 K)
while the mid-plane regions of the disk have an average temperature around
T$\sim$600 K. Our simulation also suggests that regions closer than 2
mas (10 AU) to the central source are dust-free because processes, such as
the stellar radiation and/or self-heating of the disk, sublimate the
dust at these scales \citep[see, e.g.,][]{Vaidya_2009}.

\begin{figure}[htp]
\centering
\includegraphics[width=8 cm]{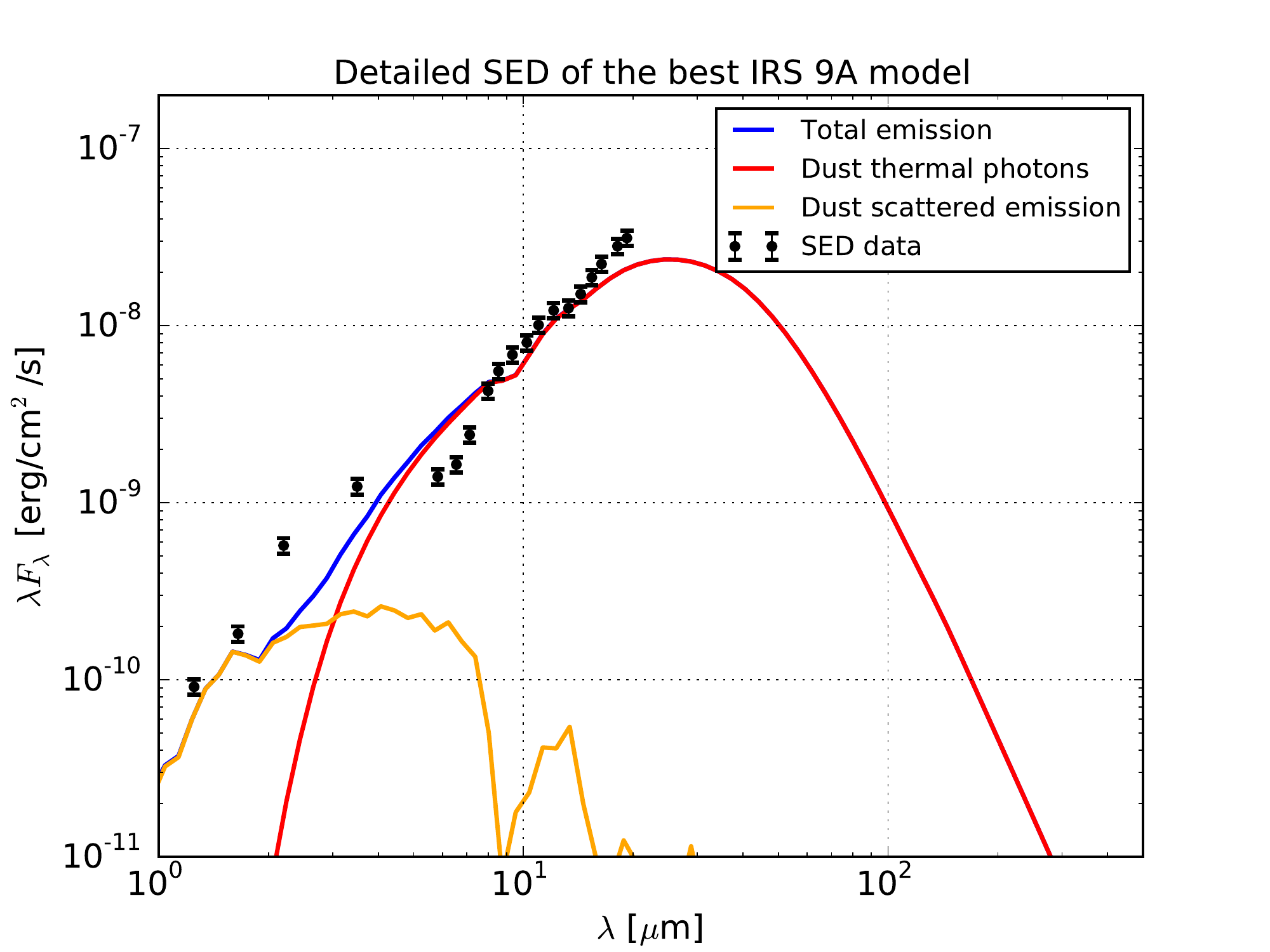} 
\caption{Spectral energy distribution of the best-fit radiative transfer
  model (blue line). The thermal and scattering contributions are displayed
  separately in red and yellow linetype respectively. }
 \label{fig:SED_detailed}
\end{figure}

With regard to the origin of the emission that is observed in the IRS\,9A SED,
Figure\,\ref{fig:SED_detailed} shows the thermal and scattering
contributions from the best-fit physical model. This plot indicates that
scattered photons from the stellar source are the most important
source of radiation at the $K_s$-filter, and that they also contribute to the
emission observed at the $L'$-filter, while thermal emission from the dust
dominates the mid- and far-infrared.

\section{Conclusions \label{sec:conclusions}}

We summarise our conclusions as follows:

a) The analytical model of a Class I YSO, described by
\citet{Whitney_2003}, appears to be a relatively good approximation of
IRS\,9A morphology. This source appears to be an embedded object that
is surrounded by a thick
envelope, which dominates the spectral energy distribution, and with 
a flat plausibly disk-like structure at its center. 
This result supports contemporary thinking in star formation, which suggest that massive stars
gain mass via accretion disks that shield part of
the infalling material from the strong radiation pressure. 

b) From our NACO/SAM data, we have confirmed the presence of a compact
, possibly disk-like, structure with an angular size $\le$30 mas. From our radiative transfer
simulations, we have found that this structure is responsible for
most of the NIR flux distribution of the IRS\,9A SED. 

c) From our best-fit radiative transfer model, we have found that the large
scale  MIR emission is dominated by the heated dust within the envelope cavity.
This is particularly supported by the observed morphology at 11.7 $\mu$m with T-ReCS. 
The model envelope has an angular size of $\sim$ 1''. 
Owing to the high luminosity of the central source,  the hot inner
regions of the envelope also emit at NIR wavelengths. 

d) The observed extended emission in the $L'-$filter NACO
image that was presented by \citet{Nurnberger_2008} is the most plausible
reasoning for the over-resolved emission observed in
the $K_s$ and $L'$ squared visibility signals.

e) Our best physical model suggests that the system of disk+envelope
is inclined $\sim$60$^{\circ}$ out of the plane of the sky, where
0$^{\circ}$ corresponds to a face-on orientation of the disk. Moreover, from our simulations, we find that smaller
inclination angles generate a large bump at near-IR wavelengths and
$V^2$ close to unity. Higher inclinations, as
suggested in prior literature models, produce low $V^2$ (failing to
reproduce the data). A fit of our model to the T-ReCS V$^2$
  and CPs allowed us to constraint $\phi \sim$120$^{\circ}$, a value
  close to the 105$^{\circ}$ (geometric Ring + Gaussian) previously reported by \citet{Vehoff_2010}.

f) From the Br$\gamma$ spectroastrometric signal, we have found that
the core of IRS\,9A is complex with ionized gas arising from different
regions (with sizes of $\sim$10 mas or $\sim$60 AU) of the morphology. New optimized spectroastrometric observations and interferometric observations
with GRAVITY/VLTI and/or MATISSE/VLTI, combined with radiative transfer emission line
models, may be useful in confirming this hypothesis and search for
the presence of additional stellar companions at the core of IRS\,9A.

g) A complete self-consistent physical scenario to describe IRS\,9A's
complex morphology is challenging, in particular fitting both the
spectral energy distribution and both small-/large-scale spatial structure.
However, systems such as this one represent important and rare test cases 
with which to confront theoretical models. In this work, we have demonstrated that a
multi-wavelength approach is necessary to unveil the physical and
geometrical properties of MYSOs. Our results indicate that optical interferometry and
spectroastrometry are important observing techniques to map the
morphological properties at the core of these targets, where important
physical phenomena occur. 

h) Future work will require more data with
optical interferometry and spectroastrometry to refine the existing models. Moreover, additional data 
at longer wavelengths (e.g., observations with ALMA) are also
necessary to better constrain the Rayleigh-Jeans part
of the SED, and to obtain more accurate information of the size and
density of the envelope. 
Similar analysis should be extended to other MYSOs to systematically  
study their properties, driving further incisive testing of massive star
formation scenarios.

%_____________________________________

\begin{acknowledgements}
We thank the referee for his/her useful comments. J.S.B., R.S. and A.A. acknowledge support by grants AYA2010-17631 and AYA2012-38491-CO2-02 of the
Spanish Ministry of Economy and Competitiveness (MINECO) cofounded
with FEDER funds, and by grant P08-TIC-4075 of the Junta de
Andaluc\'ia. R.S. acknowledges support by the Ram\'on y Cajal program
of the Spanish Ministry of Economy and
Competitiveness. J.S.B. acknowledges support by the ``JAE-PreDoc'' program
of the Spanish Consejo Superior de Investigaciones Cient\'ificas
(CSIC) and to the ESO Studentship program. This work was partly supported by OPTICON, which is supported by the European 
Commission's FP7 Capacities programme (Grant number 312430). J.S.B. thanks
R. Galv\'an-Madrid and H.-U. K\"aufl for their useful comments on this work.
\end{acknowledgements}
\bibliography{/Users/Donut/Documents/Papers/Paper_lib}

\end{document}